\documentclass[twocolumn,amsmath,amssymb,nofootinbib,eqsecnum,floatfix,prb]{revtex4}

\usepackage{graphicx}
\usepackage{dcolumn} 
\usepackage{bm, bbm}      
\usepackage{curves}
\usepackage{epic}
\usepackage{wasysym}
\usepackage{epsf, times}
\usepackage{subfigure}

\begin{document}


\title{ 
High-Temperature Criticality in Strongly Constrained Quantum Systems
      }

\author{
Claudio Castelnovo,$^1$
Claudio Chamon,$^1$
Christopher Mudry,$^2$
and
Pierre Pujol,$^3$
       }
\affiliation{
$^1$ Physics Department, Boston University, Boston, MA 02215, USA\\
$^2$
Paul Scherrer Institut, CH-5232 Villigen PSI, Switzerland\\
$^3$ Laboratoire de Physique de
l'{\'E}cole Normale Sup{\'e}rieure, Lyon, France
            }

%
%

\date{\today} 

\begin{abstract}
  The exotic nature of many strongly correlated materials at reasonably high
  temperatures, for instance cuprate superconductors in their normal state,
  has lead to the suggestion that such behavior occurs within a quantum
  critical region where the physics is controlled by the influence of a phase
  transition down at zero temperature. Such a scenario can be thought of as a
  bottom-up approach, with the zero temperature mechanisms finding a way to
  manifest critical behavior at high temperatures. Here we propose an
  alternative, top-down, mechanism by which strong kinematic constraints that
  can only be broken at extremely high temperatures are responsible for
  critical behavior at intermediate but still high temperatures. This
  critical behavior may extend all the way down to zero temperature, but this
  outcome is not one of necessity, and the system may order at low
  temperatures. We provide explicit examples of such high-temperature
  criticality when additional strong interactions are introduced in quantum
  Heisenberg, transverse field Ising, and some bosonic lattice models.
\end{abstract}

\maketitle
%
%

\section{\label{sec: Introduction}
Introduction
        }
Strongly correlated systems display very rich phase diagrams upon
varying thermodynamic parameters such as temperature, pressure, doping
concentration, magnetic field, and so on.\cite{Dagotto2005} 
At sufficiently low
temperatures, distinct phases of matter appear that are characterized
by (sometimes coexisting) long-range order of, say, the magnetic,
charge, orbital, or superconducting type. When transitions between
different zero temperature quantum phases are continuous, 
fluctuations occur at all length scales and lead to power law behavior for
correlation functions of the order parameter at the critical
coupling. Tuning the temperature slightly away from zero decreases the
strength of the order parameter fluctuations but as long as they
remain sufficiently strong they lead to scaling laws that are
insensitive to the microscopic details and, to a large degree,
universality has emerged.

The temperature range for which scaling laws apply is a measure of how strong
fluctuations are. In strongly correlated systems such as organic materials,
high-$T^{\ }_c$ superconductors, etc, these scaling laws extend to
surprisingly high temperatures above the critical temperature. This fact is
attributed to the presence of very strong one- or two-dimensional
fluctuations of the order parameter that are predominantly quantum at
sufficiently short length scales. How large the temperature is at which
quantum fluctuations are effectively observed is a matter of intense
debate.~\cite{Kopp-Chakravarty} For example it has been proposed that many
exotic properties in the so-called pseudogap regime of high-$T^{\
}_{\mathrm{c}}$ superconductors originate in a hidden quantum critical
point.\cite{Varma1997,Sachdev03,Kivelson03} In this scenario, increasing the
doping concentration induces at zero temperature a continuous phase
transition between two different states of matter below the superconducting
dome, that is reflected in the strange metallic properties of the normal state
above the superconducting dome.
\begin{figure}[!hb]
\center
\includegraphics[width=0.95\columnwidth]{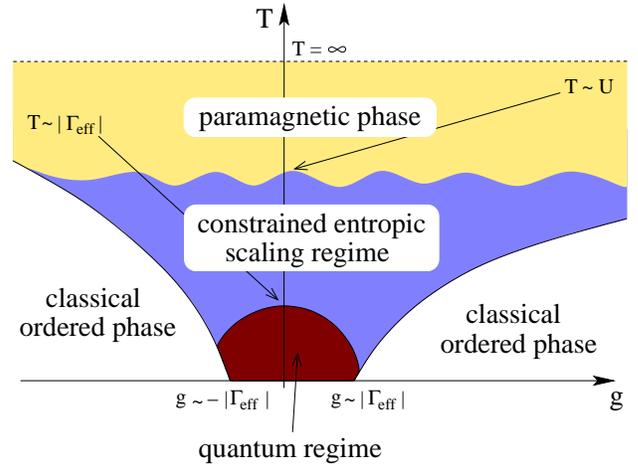}
\caption{
\label{fig: general ph diag} 
(Color Online) -- 
Generic phase diagram for a strongly constrained quantum system that
satisfies conditions 1) to 6) in Sec.~\ref{sec: Introduction}.
The parameter space encodes the competition between two energy scales,
the temperature $T$ and a characteristic energy scale
$g$ that selects a classical ordered phase, 
while $\Gamma^{\ }_{\mathrm{eff}}$ and $U$ are held fixed. 
A quantum critical scaling regime, if it exists, is restricted to
a region represented by the upper half of a disk of radius
$\sim |\Gamma^{\ }_{\mathrm{eff}}|$ and centered at the origin
$(g,T)=(0,0)$ of parameter space. 
The focus of this paper is on the \textit{constrained entropic scaling regime}
at temperatures intermediate between the small characteristic quantum
energy scale $\Gamma^{\ }_{\mathrm{eff}}$ and the large
characteristic energy scale $U$ set by a strong constraint. 
At a fixed temperature, the constrained entropic scaling regime
terminates in a phase transition that needs not be continuous
upon increasing $|g|$. At fixed $g$, the constrained entropic scaling regime
crosses over to the conventional high-temperature phase, say a paramagnetic
one for spin degrees of freedom, when $T$ is of the order of $U$.
The transition from the constrained entropic scaling regime
to the quantum regime upon lowering $T$ at fixed $g$ is system specific.
        }
\end{figure}

In this paper, we present an alternative picture where the critical
behavior in some strongly interacting systems is present at high
temperatures due to strong kinematic constraints in the quantum
Hamiltonian. The system remains critical as long as the constraints
are respected, i.e., below some large energy scale, corresponding 
to the largest coupling in the Hamiltonian. This critical phase
becomes the physically important, universal feature from which a
scaling regime can descend.  The zero-temperature physics is instead
system-specific, with a rich variety of different ordered phases, and
may or may not allow for the finite temperature criticality to survive
all the way to zero-temperature. We also discuss how this picture
brings about four distinct scenarios for the finite temperature
behavior of the system, depending on the type of interactions present
in the quantum Hamiltonian. In particular, we show how one scenario
naturally leads to a quantum system exhibiting an exotic correlation
length that increases with increasing temperature, over a wide range
of temperatures. 

The making of a system displaying high-temperature criticality is
as follows.

\vspace{\baselineskip}
\noindent
1) At least one characteristic energy scale in the quantum Hamiltonian, 
   say $U$, is much larger than all others. Thus, $U$ splits the
   Hilbert space in sectors separated by large energy gaps. In particular, 
   the infinite $U$ limit projects the Hilbert space onto the space
   $\mathcal{H}^{\ }_{0,U}$ of allowed states.

\vspace{\baselineskip}
\noindent
2) The quantum dynamics is generated by the terms that do not commute
   with the $U$-term in the quantum Hamiltonian. 
   However, one needs a process of order $n\ge 2$ in 
   the perturbative expansion 
   in the characteristic energy scale of these terms,
   $\Gamma$, to generate an effective quantum coupling between
   allowed states.
   Hence, this effective quantum coupling 
   $\Gamma^{\ }_{\mathrm{eff}}\sim \Gamma (\Gamma/U)^{n-1}$ 
   can be very small, $|\Gamma^{\ }_{\mathrm{eff}}| \ll |\Gamma|$.

\vspace{\baselineskip}
\noindent
3) By squashing down the effective quantum coupling  
   $\Gamma^{\ }_{\mathrm{eff}}$,
   one opens a hierarchy of temperature scales, which we discuss in
   detail in the paper. In particular, already at reasonably low
   temperatures (above the small $|\Gamma^{\ }_{\mathrm{eff}}|$), 
   the effective quantum term can be neglected in the calculation of 
   equilibrium thermodynamic quantities.

\vspace{\baselineskip}
\noindent
4) Although the equilibrium \textit{thermodynamics} for 
   $T\gg |\Gamma^{\ }_{\mathrm{eff}}|$ 
   is classical, the \textit{dynamics} is still
   quantum. The reason is the following. If the system were to rely on
   thermally activated processes to move within the restricted Hilbert
   space $\mathcal{H}^{\ }_{0,U}$, the characteristic time-scale would
   be, for a small system-bath coupling $\gamma > 0$, $\tau^{\ }_{T}=
   \gamma^{-1}\exp(U/T)$, which is astronomical for temperatures well
   below the large energy scale $U$. There are also virtual processes
   due to system-bath coupling that bypass thermal activation, but
   whose effective coupling is suppressed: $\gamma^{\ }_{\mathrm{eff}}
   \sim \gamma (\gamma/U)^{n-1}\ll\gamma$, much alike the intrinsic
   terms that do not commute with the $U$-term.  The waiting time for
   a quantum tunneling event is $\tau^{\ }_{Q}={\rm
   min}(|\Gamma^{-1}_{\mathrm{eff}}|,\gamma^{-1}_{\mathrm{eff}})$, which
   depends only algebraically on $U$. Since $\gamma\ll |\Gamma|$ (the
   bath coupling should be the weakest term in order not to perturb 
   the energy levels of the system), the intrinsic quantum
   dynamics provides the smallest dynamical time scale. Hence, a
   phantom of quantum mechanics in the form of sporadic tunneling
   events between which coherence is lost provides the fastest
   mechanism for the system to reach classical thermodynamic
   equilibrium when $|\Gamma^{\ }_{\mathrm{eff}}|\ll T\ll U$.

\vspace{\baselineskip}
\noindent
5) The strong constraint on the allowed states 
   imposed by taking the limit $U\to\infty$ first makes the
   system critical in the constrained entropic limit $T\to\infty$ 
   with all the remaining characteristic energy scales held fixed.
   We call such a critical point
   \textit{a constrained entropic critical point}. 
   The system's properties at low 
   temperatures compared to $U$ are controlled by the close
   proximity to this purely constrained entropic critical point.
   Ice, coloring, and dimer models provide 
   examples of constrained entropic critical points in systems with hard
   constraints. 

\vspace{\baselineskip}
\noindent
6) Finally, if there are other terms in the Hamiltonian with 
   characteristic energy scale $g$, $|g|\ll U$, but that commute with the
   $U$-term, there can be different phases and transitions among them
   when $T/|g|$ is ${\cal O}(1)$.  When $U \gg T\gg |g|$, the physics is 
   controlled by the proximity to the constrained entropic critical point. 
   Even though the constrained entropic critical point may be unstable, the
   nearby renormalization group (RG) trajectories feel its presence until
   the RG scale $T\sim U$ is reached, beyond which the constraint becomes
   immaterial.  The featureless unconstrained (stable) fixed point takes
   over at that stage.

The phase diagram in Fig.~\ref{fig: general ph diag} 
exemplifies the high-temperature critical behavior arising from kinematic
constraints that we highlight in this paper. 
It shows a large region 
-- the constrained entropic critical regime -- 
standing in between two ordered phases. 
The constrained entropic critical region exists because of the
strong constraints imposed by the large energy scale $U$ in the
problem, and it covers a high-temperature range that ends at the
extreme limit of temperatures of order $U$, where the system becomes
featureless in that it is controlled by its proximity to
the unconstrained entropic fixed point (paramagnetic phase). 
The ordered phases to the left or right of
the constrained entropic critical regime 
have a classical origin in Fig.~\ref{fig: general ph diag}, 
since there the second largest energy scale $|g|$ 
is associated to an operator that commutes with the constraint. 
If we fix $\Gamma^{\ }_{\mathrm{eff}}$ and vary $g$, as in 
Fig.~\ref{fig: general ph diag}, 
the high-temperature constrained entropic critical region 
sits on top of a quantum phase with radius of the order of 
$|\Gamma^{\ }_{\mathrm{eff}}|$ around the origin $T=g=0$. 
Quantum criticality may exist only within this small region.

We illustrate this constraint-based critical behavior at high
temperatures using simple case studies, a constrained quantum
Heisenberg model and a constrained Ising model in a transverse field
both on the honeycomb lattice, and a constrained bosonic model on the
square lattice that leads to the quantum dimer model at low
energies. The single-band Hubbard model, where a large onsite
repulsive term $U$ is the dominant energy scale, fails to fulfill
condition 5), and correlations decay exponentially fast with
separation beyond a characteristic length scale of the order of the
lattice spacing. However, this situation may change if one considers
extended Hubbard models with large nearest-neighbor coupling $V$, next
nearest-neighbor coupling $V'$, etc, at commensurate fillings. We
shall comment on this situation in the conclusions, and discuss, for
instance, the possible connection between the ideas of constrained
entropic criticality and those of fluctuating
stripes.~\cite{Kivelson03} In addition, power-law behavior at high
temperatures was also shown to occur in the context of locally
fluctuating bond currents in d-density wave states by Chakravarty in
Ref.~\onlinecite{Chakravarty02}, who also recognized the importance of
the constraints in determining the long distance correlations in the
system.

The plan of the paper is as follows. Four examples of constrained quantum
models are introduced in Sec.~\ref{sec: four examples}. The regimes of
temperature that are of relevance to this paper are presented in
Sec.~\ref{sec: Competing characteristic energy scales}. The constrained
entropic scaling regime is described in Sec.~\ref{sec: Constrained entropic
scaling regime}.  Realizations of constrained entropic scaling regimes for
quantum XXZ Heisenberg and transverse field Ising models on the one hand, and
lattice bosonic models on the other hand, can be found in Sec.~\ref{sec:
constrained Ising} and Sec.~\ref{sec: The square lattice dimer model},
respectively. In Sec.~\ref{sec: Quantum-discussion} we discuss the regime
when the quantum scale $|\Gamma^{\ }_{\mathrm{eff}}|$ is the largest scale
below $U$, i.e., $|g|\ll |\Gamma^{\ }_{\mathrm{eff}}|\ll U$, 
and we conclude in Sec.~\ref{sec: Conclusions}.
%
%

\section{\label{sec: four examples}
Four examples of constrained quantum models
        }
Constrained quantum mechanical systems are described by Hamiltonians
of the generic form
\begin{subequations}
\begin{eqnarray}
&&
\hat{H}= 
\hat{H}^{\ }_{g,\Gamma} 
+ 
\hat{H}^{\ }_{U},
\\
&&
\hat{H}^{\ }_{g,\Gamma}=
\hat{H}^{\ }_{g} 
+
\hat{H}^{\ }_{\Gamma},
\end{eqnarray}
where it is assumed that 
$\hat{H}^{\ }_{U}$ can be diagonalized in some preferred basis $\mathcal{B}$
that spans the Hilbert space $\mathcal{H}$ on which $\hat{H}$ is defined,
$\hat{H}^{\ }_{g}$ commutes with  $\hat{H}^{\ }_{U}$,
\begin{eqnarray}
[\hat{H}^{\ }_{g},\hat{H}^{\ }_{U}]=0,
\end{eqnarray}
$\hat{H}^{\ }_{\Gamma}$ does not commute with  $H^{\ }_{U}$
\begin{eqnarray}
[\hat{H}^{\ }_{\Gamma},\hat{H}^{\ }_{U}]\neq0,
\end{eqnarray}
and the characteristic energy scale $U$ of $\hat{H}^{\ }_{U}$ is much larger
than the characteristic energy scales $g$ and $\Gamma$ of
$\hat{H}^{\ }_{g}$
and
$\hat{H}^{\ }_{\Gamma}$,
respectively,
\begin{eqnarray}
|g|,|\Gamma|\ll U.
\end{eqnarray}
\end{subequations}

A famous example of a constrained many-body Hamiltonian is 
the single-band Hubbard model~\cite{Hubbard1963} 
\begin{subequations}
\label{eq: example 1}
\begin{eqnarray}
\hat{H}&:=&
-
\mu
\sum_{i}
\sum_{\sigma=\uparrow,\downarrow}
\hat{c}^{\dag}_{i\sigma}
\hat{c}^{\   }_{i\sigma}
\label{eq: example 1 a}
\\
&&
-
t
\sum_{\langle ij\rangle}
\sum_{\sigma=\uparrow,\downarrow}
\left(
\hat{c}^{\dag}_{i\sigma}
\hat{c}^{\   }_{j\sigma}
+
\hat{c}^{\dag}_{j\sigma}
\hat{c}^{\   }_{i\sigma}
\right)
\label{eq: example 1 b}
\\
&&
+
U
\sum_{i}
\prod_{\sigma=\uparrow,\downarrow}
\left(
\hat{c}^{\dag}_{i\sigma}
\hat{c}^{\   }_{i\sigma}
\right)
\label{eq: example 1 c}
\end{eqnarray}
\end{subequations}
that acts on the fermionic Fock space generated by the creation 
$\hat{c}^{\dag}_{i\sigma}$ and annihilation $\hat{c}^{\   }_{i\sigma}$
operators for electrons carrying the site index $i$ and the
spin index $\sigma$. The preferred basis $\mathcal{B}$ is the basis specified 
by all the local fermionic occupation numbers 
$
\hat{c}^{\dag}_{i\sigma}
\hat{c}^{\   }_{i\sigma}
$
where $i$ runs over all the lattice sites and $\sigma$ over 
the spin up or down. 
In this preferred basis,
the chemical potential~(\ref{eq: example 1 a})
and the potential energy~(\ref{eq: example 1 c})
are diagonal whereas the kinetic energy~(\ref{eq: example 1 b}) is not.
The chemical potential $\mu$ thus plays the role of $g$ while
the hopping amplitude $t$ plays the role of $\Gamma$.

A second example is the constrained XXZ quantum spin-1/2 magnet 
defined on the two-dimensional honeycomb lattice by the 
Hamiltonian~\cite{Muller1982,Castelnovo2005}
\begin{subequations}
\label{eq: example 2}
\begin{eqnarray}
\hat{H}&:=&
-
J
\sum_{\langle ij\rangle}
\hat{\sigma}^{\mathrm{z}}_{i}
\hat{\sigma}^{\mathrm{z}}_{j}
-
\sum_{i=1}^{N}
\left[
h
+
(-1)^i
h^{\ }_{s}
\right] 
\hat{\sigma}^{\rm z}_{i} 
\label{eq: example 2 a}
\\
&&
-
\Gamma
\sum_{\langle ij\rangle}
\left(
\hat{\sigma}^{\mathrm{x}}_{i}
\hat{\sigma}^{\mathrm{x}}_{j}
+
\hat{\sigma}^{\mathrm{y}}_{i}
\hat{\sigma}^{\mathrm{y}}_{j}
\right)
\label{eq: example 2 b}
\\
&&
+
U 
\sum_{\hexagon} 
\left[ 
1 
- 
\cos 
\left(
2\pi \sum_{i\in\hexagon} \hat{\sigma}^{\rm z}_{i}/3
\right)
\right]
\label{eq: example 2 c}
\end{eqnarray}
\end{subequations}
where the sum over the symbol $\hexagon$ is to be understood as a summation
over all elementary hexagons making up the honeycomb lattice and the
three $2\times2$ Pauli matrices are denoted by 
$\hat{\sigma}^{\mathrm{x}}$, $\hat{\sigma}^{\mathrm{y}}$, 
$\hat{\sigma}^{\mathrm{z}}$, 
respectively. The preferred basis $\mathcal{B}$ is the basis specified by all 
the eigenstates with eigenvalues $\sigma^{\mathrm{z}}_{i}$ of
$\hat{\sigma}^{\mathrm{z}}_{i}$, where $i$ runs over all $N$ sites
of the honeycomb lattice. In this preferred basis
the potential energy~(\ref{eq: example 2 c})
and the longitudinal one- and two-body interaction~(\ref{eq: example 2 a})
are diagonal whereas the transverse two-body 
interaction~(\ref{eq: example 2 b}) is not.
The exchange coupling $J$, the uniform magnetic field $h$, and the staggered 
magnetic field $h^{\ }_s$ thus play the role of three different 
diagonal couplings $g$. 

A third example is the constrained quantum Ising model in a transverse 
field~\cite{Elliott1970,Castelnovo2005} 
\begin{subequations}
\label{eq: example 3}
\begin{eqnarray}
\hat{H}&:=&
-
J
\sum_{\langle ij\rangle}
\hat{\sigma}^{\mathrm{z}}_{i}
\hat{\sigma}^{\mathrm{z}}_{j}
-
\sum_{i=1}^{N}
\left[
h
+
(-1)^i
h^{\ }_{s}
\right] 
\hat{\sigma}^{\rm z}_{i} 
\label{eq: example 3 a}
\\
&&
-
\Gamma
\sum_{i=1}^{N}
\hat{\sigma}^{\mathrm{x}}_{i}
\label{eq: example 3 b}
\\
&&
+
U 
\sum_{\hexagon} 
\left[ 
1 
- 
\cos 
\left(
2\pi \sum_{i\in\hexagon} \hat{\sigma}^{\rm z}_{i}/3
\right)
\right]
\label{eq: example 3 c}
\end{eqnarray}
\end{subequations}
that shares with the spin-1/2 quantum
Hamiltonian~(\ref{eq: example 2})
the same preferred basis. 
In this preferred basis, 
the potential energy~(\ref{eq: example 3 c})
and the one- and two-body interactions~(\ref{eq: example 3 a})
are diagonal whereas the one-body transverse field~(\ref{eq: example 3 b})
is not. 
The exchange coupling $J$, the uniform magnetic field $h$, and the staggered 
magnetic field $h^{\ }_s$ thus play again the role of three different 
couplings $g$. 

Our last example is the quantum Hamiltonian 
\begin{subequations}
\label{eq: example 4}
\begin{eqnarray}
\hat{H}&:=&
v
\sum_{i=1}^{N}
\Big[
\left(
\hat{b}^{\dag}_{i,i+\boldsymbol{x}}
\hat{b}^{\   }_{i,i+\boldsymbol{x}}
\right)
\left(
\hat{b}
^{\dag}_{i+\boldsymbol{y},i+\boldsymbol{y}+\boldsymbol{x}}
\hat{b}
^{\   }_{i+\boldsymbol{y},i+\boldsymbol{y}+\boldsymbol{x}}
\right)
\nonumber\\
&&
\hphantom{v\sum_{i=1}^{N}}
+
\boldsymbol{x}
\leftrightarrow
\boldsymbol{y}
\Big]
\label{eq: example 4 a}
\\
&&
-
t
\sum_{i=1}^{N}
\left[
\hat{b}^{\dag}_{i,i+\boldsymbol{x}}
\left(
\hat{b}^{\   }_{i,i+\boldsymbol{y}}
+
\hat{b}^{\   }_{i+\boldsymbol{x},i+\boldsymbol{x}+\boldsymbol{y}}
\right.
\right.
\nonumber\\
&&
\qquad\qquad\;\;\;\;\; 
\left.
\left.
+
\hat{b}^{\   }_{i-\boldsymbol{y},i}
+
\hat{b}^{\   }_{i+\boldsymbol{x}-\boldsymbol{y},i+\boldsymbol{x}}
\right)
+
\mathrm{H.c.}
\vphantom{\sum}\right]
\nonumber\\
&&
\label{eq: example 4 b}
\\
&&
+
U 
\sum_{i=1}^{N} 
\left[
1 
-
\!\!\sum_{\boldsymbol{e}=\boldsymbol{x},\boldsymbol{y}}\!\!
\left(
\hat{b}^{\dag}_{i-\boldsymbol{e},i}
\hat{b}^{\   }_{i-\boldsymbol{e},i}
+
\hat{b}^{\dag}_{i,i+\boldsymbol{e}}
\hat{b}^{\   }_{i,i+\boldsymbol{e}}
\right)
\right]^{2}
\nonumber\\
&&
\label{eq: example 4 c}
\end{eqnarray}
\end{subequations}
that acts on the bosonic Fock space generated by the
bosonic creation $\hat{b}^{\dag}_{i,i+\boldsymbol{e}}$
and annihilation $\hat{b}^{\   }_{i,i+\boldsymbol{e}}$
operators defined on the mid-points of the nearest-neighbor
links of the square lattice.
The preferred basis $\mathcal{B}$ is the basis specified by all the local
bosonic occupation numbers, i.e., the eigenvalues
$n^{\   }_{i,i+\boldsymbol{e}}$ of
\begin{eqnarray}
\hat{n}^{\   }_{i,i+\boldsymbol{e}}:=
\hat{b}^{\dag}_{i,i+\boldsymbol{e}}
\hat{b}^{\   }_{i,i+\boldsymbol{e}}
\end{eqnarray}
for all the $2N$ links $i,i+\boldsymbol{e}$
where $i$ runs over the $N$ sites of the square lattice
and $\boldsymbol{e}$ over its two generating vectors
$\boldsymbol{x}$ 
and  
$\boldsymbol{y}$.
In this preferred basis
the potential energy~(\ref{eq: example 4 c})
and the four-body interaction~(\ref{eq: example 4 a})
are diagonal whereas the kinetic energy~(\ref{eq: example 4 b})
is not. The coupling $v$ thus plays the role of $g$ while
the coupling $t$ plays the role of $\Gamma$.

By assumption, the contribution $\hat{H}^{\ }_{U}$ encodes the largest 
characteristic energy scale $U$ in the problem.
As long as all other energy scales are much smaller than $U$, 
the low-energy physics is captured by an effective Hamiltonian
$\hat{H}^{\ }_{\mathrm{eff}}$
that is defined on the Hilbert space restricted to the lowest energy 
eigenstates of $\hat{H}^{\ }_{U}$. 
$\hat{H}^{\ }_{\mathrm{eff}}$ can be systematically deduced from $\hat{H}$ 
by treating all the contributions $\hat{H}^{\ }_{\Gamma}$
to $\hat{H}^{\ }_{g,\Gamma}$ that do not commute with
$\hat{H}^{\ }_{U}$ within perturbation theory.
This effective model is particularly interesting whenever the ground state 
manifold $\mathcal{H}_{0,U}$ of
$\hat{H}^{\ }_{U}$ is extensively degenerate, as is the case for all 
examples~(\ref{eq: example 1}-\ref{eq: example 4}). 
Within degenerate perturbation theory,
$\hat{H}^{\ }_{\mathrm{eff}}$ is given by
\begin{subequations}
\label{eq: H eff}
\begin{eqnarray}
\label{eq: effective Hamiltonian} 
\hat{H}^{\ }_{\mathrm{eff}}= 
\hat{H}^{\ }_{g} 
- 
\Gamma^{\ }_{\mathrm{eff}} 
\hat{H}^{(n)}_{\Gamma/U} 
\end{eqnarray}
with
\begin{eqnarray}
\label{eq: effective Gamma} 
\Gamma^{\ }_{\mathrm{eff}} \propto
\frac{\Gamma^{n}}{U^{n-1}}
\end{eqnarray}
\end{subequations}
and
$\hat{H}^{(n)}_{\Gamma/U}$ of order zero in $\Gamma/U$,
whereby it is understood that 
$\hat{H}^{\ }_{\mathrm{eff}}$
acts only on the subspace 
$\mathcal{H}^{\ }_{0,U}$ of the unconstrained
Hilbert space with basis $\mathcal{B}$.
The order $n$ 
and the form taken by $\hat{H}^{(n)}_{\Gamma/U}$
in Eq.~(\ref{eq: effective Hamiltonian})
are model dependent. 

For the Hubbard model~(\ref{eq: example 1}), 
$n=1$ and\cite{Anderson1963,Emery1976,Chao1977} 
\begin{eqnarray}
\hat{H}^{\ }_{\mathrm{eff}} &=&
-
\mu
\sum_{i}
\sum_{\sigma=\uparrow,\downarrow}
\hat{c}^{\dag}_{i\sigma}
\hat{c}^{\   }_{i\sigma}
\nonumber\\
&&
-
t
\left\{
  \sum_{\langle ij\rangle}
  \sum_{\sigma=\uparrow,\downarrow}
  \left(
    \hat{c}^{\dag}_{i\sigma}
    \hat{c}^{\   }_{j\sigma}
    +
    \hat{c}^{\dag}_{j\sigma}
    \hat{c}^{\   }_{i\sigma}
  \right)
\right.
\\
&& 
\left.
\qquad
+ 
\frac{4t}{U}
  \sum_{\langle ij\rangle}
  \left(
    \hat{\boldsymbol{S}}^{\ }_{i}
    \cdot
    \hat{\boldsymbol{S}}^{\ }_{j}
    - 
    \frac{1}{4}
    \hat{n}^{\ }_{i}
    \hat{n}^{\ }_{j} 
  \right)
  + 
  \cdots
\vphantom{\sum_{\langle ij\rangle}}
\right\}
\nonumber 
\end{eqnarray}
acts on the subspace $\mathcal{H}^{\ }_{0,U}$
of the fermionic Fock space with no more than one electron per site. 
Here, we have introduced the fermionic bilinears
$
\hat{\boldsymbol{S}}^{\ }_{i}:=
      \hat{c}^{\dag}_{i\sigma}
  \frac{\boldsymbol{\sigma}_{\sigma\sigma^{\prime}}}{2} 
      \hat{c}^{\   }_{i\sigma^{\prime}}
$
and
$
\hat{n}^{\ }_{i}:=
      \hat{c}^{\dag}_{i\sigma}
      \hat{c}^{\   }_{i\sigma}
$
(summation over repeated spin indices is implied).
Notice that, at half-filling,
all contributions to first order in $t$ vanish 
so that $\Gamma_{\textrm{eff}}$ becomes second order in $t$ ($n=2$). 

For the constrained XXZ quantum spin-1/2 magnet on the honeycomb lattice%
~(\ref{eq: example 2}),
$n=3$ in Eqs.~(\ref{eq: H eff}) 
whereby the constrained Hilbert space
is the subspace $\mathcal{H}^{\ }_{0,U}$
defined by all states of the form
$|\sigma^{\rm z}_{1},\ldots,\sigma^{\rm z}_{N}\rangle$
such that the magnetization of each elementary hexagonal plaquette is 
some integer multiple of 3, i.e., 
\begin{eqnarray}
\sum_{i\in\hexagon}\sigma^{\rm z}_{i}/3\in\mathbb{Z}. 
\end{eqnarray}
For the constrained quantum Ising model in a transverse field 
on the honeycomb lattice%
~(\ref{eq: example 3}),
$n=6$, and the constrained Hilbert space is the same as in example 
~(\ref{eq: example 2}).

At last, $n=2$ for the bosonic Hamiltonian%
~(\ref{eq: example 4})
whereby the constrained Hilbert space
is the subspace $\mathcal{H}^{\ }_{0,U}$
spanned by all states of the form
$|\ldots,n^{\ }_{\langle ij\rangle},\ldots\rangle$, 
where $\langle ij\rangle$ runs over all $2N$ links of the
square lattice made of $N$ sites, and the boson occupation numbers 
are restricted to
\label{eq: dimer condition}
\begin{eqnarray}
1=
\sum_{\boldsymbol{e}=\boldsymbol{x},\boldsymbol{y}}\!\!
\left(
n^{\ }_{i-\boldsymbol{e},i}
+
n^{\ }_{i,i+\boldsymbol{e}}
\right)
\end{eqnarray}
for all $N$ sites $i$. This condition is satisfied if 
$n^{\ }_{\langle ij\rangle}=0,1$ and only one link out of the four 
connected to each vertex is occupied by a boson. This constrained system 
is equivalent to a square lattice quantum dimer model. 

The last three models will be studied in detail in 
Sec.~\ref{sec: constrained Ising} 
and Sec.~\ref{sec: The square lattice dimer model}, 
which the reader is referred to for an example-based approach. 
In Sec.~\ref{sec: Competing characteristic energy scales} 
and Sec.~\ref{sec: Constrained entropic scaling regime}
we now discuss 
the possible topologies of the phase diagram
when a strongly constrained quantum system
possesses a constrained critical regime at intermediary temperatures.
%
%

\section{\label{sec: Competing characteristic energy scales}
Competing characteristic energy scales
        } 
{}From the generic form taken by the low energy
Hamiltonian~(\ref{eq: H eff})
acting on the ground state manifold $\mathcal{H}^{\ }_{0,U}$
and from the assumption that $|g|, |\Gamma| \ll U$, we deduce 
the existence of at least three regimes of temperatures
provided any one of the hierarchies of energy scales
\begin{subequations}
\label{eq: 3 regimes}
\begin{eqnarray}
&&
T \ll 
|\Gamma^{\ }_{\mathrm{eff}}|,|g| \ll
U, 
\label{eq: regime A}
\\
&&
|\Gamma^{\ }_{\mathrm{eff}}|,|g| \ll T \ll U, 
\label{eq: regime B}
\\
&&
|\Gamma^{\ }_{\mathrm{eff}}|,|g| \ll
U \ll 
T.
\label{eq: regime C}
\end{eqnarray}
\end{subequations}
holds. 
In addition to these three temperature regimes, strongly constrained 
systems described by the low energy Hamiltonian~(\ref{eq: H eff})  
can exhibit a fourth regime if $|g|$ is much larger than 
all effective off-diagonal couplings. 
This is the case for example when $|g| \gtrsim |\Gamma|$ and the off-diagonal 
terms in $\hat{H}^{\ }_{\Gamma}$ do not contribute to first order in $\Gamma$, 
i.e., $n>1$ in Eq.~(\ref{eq: effective Gamma}). 
If so, $|\Gamma^{\ }_{\mathrm{eff}}|$ is guaranteed to be much 
smaller than $g$ and the temperature regime~(\ref{eq: regime A}) can be 
further subdivided into two distinct ones, for a total of four regimes 
\begin{subequations}
\label{eq: 4 regimes}
\begin{eqnarray}
&&
T \ll 
|\Gamma^{\ }_{\mathrm{eff}}| \ll |g| \ll
U,
\label{eq: regime 1}
\\
&&
|\Gamma^{\ }_{\mathrm{eff}}| \ll 
T \ll |g| \ll 
U, 
\label{eq: regime 2}
\\
&&
|\Gamma^{\ }_{\mathrm{eff}}| \ll 
|g| \ll 
T 
\ll U, 
\label{eq: regime 3}
\\
&&
|\Gamma^{\ }_{\mathrm{eff}}| \ll |g| \ll
U \ll 
T.
\label{eq: regime 4}
\end{eqnarray}
\end{subequations}

For simplicity, we shall consider only two limiting cases 
that encode the competition between the classical energy scale
$g$ 
and the quantum energy scale
$\Gamma^{\ }_{\textrm{eff}}$
in this paper.
The first occurs when 
$|g|\gg|\Gamma^{\ }_{\textrm{eff}}|$ 
in the temperature regime~(\ref{eq: regime 3}).
The second occurs when 
$|g|\ll|\Gamma^{\ }_{\mathrm{eff}}|$ 
in the temperature regime~(\ref{eq: regime B}).
Both situations can be realized in
examples~(\ref{eq: example 2}-\ref{eq: example 4}). 
In particular, the case of a quantum energy scale dominating over 
the classical one is of relevance to the Hubbard model close to half-filling, 
to quantum XXZ Heisenberg magnets with strong XY exchange anisotropy,
to Ising magnets subjected to strong transverse magnetic fields,
or to constrained bosons whose kinetic energy dominates over their
interactions. We postpone the discussion of the second case to 
Sec.~\ref{sec: Quantum-discussion}, while we consider here the case 
$|g|\gg|\Gamma^{\ }_{\textrm{eff}}|$. 

The quantum world at $T=0$ is governed by a delicate
competition between the diagonal and off-diagonal energy scales.
As these characteristic energy scales are varied, 
quantum phase transitions can take place between different 
states of matter that usually support long-range order. 
Universality emerges at fine-tuned quantum critical points
where the transitions between different states 
of matter are continuous. The signature of a quantum critical point
can manifest itself as a scaling regime as long as temperatures are not 
too large, say when Eq.~(\ref{eq: regime 1}) holds.

Upon increasing the temperature from $T=0$, one leaves
the quantum regime~(\ref{eq: regime 1})
once $T$ becomes larger than
the quantum coupling $|\Gamma^{\ }_{\mathrm{eff}}|$
in the restricted Hilbert space.
Beyond the quantum regime one distinguishes the three 
regimes%
~(\ref{eq: regime 2}),%
~(\ref{eq: regime 3}), 
and%
~(\ref{eq: regime 4}). 
If the temperature is much smaller than $|g|$ 
thermal fluctuations are dominated by the classical energy 
scale $g$.
If the temperature is raised to values that are much larger than $|g|$ but 
that remain much smaller than the characteristic constraint energy $U$,
then the thermal fluctuations are predominantly entropic in character, 
with the entropy of the classical constrained phase space isomorphic 
to the preferred basis
$\mathcal{B}^{\ }_{0,U}$ that spans the ground-state manifold
$\mathcal{H}^{\ }_{0,U}$ of $\hat{H}^{\ }_{U}$.
Finally, once the temperature becomes the largest energy scale in the 
problem, the thermal fluctuations are still entropic in character, but 
now with the entropy of the classical unconstrained phase space isomorphic
to the basis $\mathcal{B}$ of $\hat{H}$. 
The transitions between these regimes can take place 
through phase transitions or through crossovers. 

The two entropic regimes~(\ref{eq: regime 3}) and~(\ref{eq: regime 4})
do not always need to be qualitatively different. This is the case for the
Hubbard model for which all connected spatial correlation functions
between the local electronic densities 
$\hat{n}^{\ }_{i\sigma}=\hat{c}^{\dag}_{i\sigma}\hat{c}^{\ }_{i\sigma}$ 
in the entropic regime~(\ref{eq: regime 3}) decay in a qualitatively
similar way as in the entropic regime~(\ref{eq: regime 4}), i.e.,
exponentially fast with separation beyond a characteristic length
scale of the order of the lattice spacing. The situation may change if
one considers extended Hubbard models with nearest-neighbor
coupling $V$, next nearest-neighbor coupling $V'$, etc, 
at commensurate fillings. 
We shall comment on this situation in the conclusions.
In examples~(\ref{eq: example 2}-\ref{eq: example 4}), 
we shall show below that 
the two entropic regimes~(\ref{eq: regime 3})
and~(\ref{eq: regime 4})
are qualitatively different as measured by the temperature dependence
and order of magnitude of the correlation length characterizing the onset
of exponential decay 
in spatial correlation functions. 
%
%

\section{\label{sec: Constrained entropic scaling regime}
The constrained entropic scaling regime
        }
In this section we shall study the generic features of the 
\textit{constrained entropic regime}~(\ref{eq: regime 3}) 
for Hamiltonians of the type~(\ref{eq: example 2}-\ref{eq: example 4}). 
All the assumptions and different cases considered here will be supported 
with explicit examples in 
Sec.~\ref{sec: constrained Ising} 
and 
Sec.~\ref{sec: The square lattice dimer model}. 
The reader who may be unfamiliar with the physics of constrained models 
is referred to those two sections for an example-based approach. 

Since quantum dynamics is of no qualitative
relevance in regime~(\ref{eq: regime 3}), we shall always assume that 
$|\Gamma^{\ }_{\mathrm{eff}}|/T\to0$ while keeping the ratios 
$|g|/T\ll1$ and $T/U\ll1$ fixed.  
Were it not for the presence of the constraint in
regime~(\ref{eq: regime 3}), set by the large energy scale $U \gg T$, 
the condition $T\gg|g|$ would place the
system deep into a massive phase, i.e., 
correlation functions would decay exponentially in
space with a characteristic decay length, the correlation length, of
the order of the lattice spacing $\mathfrak{a}$. 
We are going to argue that, 
in examples~(\ref{eq: example 2}-\ref{eq: example 4}), 
the constraining energy scale $U$ induces a correlation length much 
larger than the lattice spacing and possibly increasing with temperature in
regime~(\ref{eq: regime 3})! 

Without loss of generality $g>0$ is assumed in the remainder of this section. 
%
%

\subsection{
The scaling limit $g/T,T/U\to0$
           }
We begin our analysis by considering the scaling limit
$g/T,T/U\to0$. In this limit, 
all entropic fluctuations are restricted to
the classical configuration space isomorphic to the basis
$\mathcal{B}^{\ }_{0,U}$. This hard constraint has dramatic
consequences on thermal averages in 
examples~(\ref{eq: example 2}-\ref{eq: example 4}).
Indeed any spatial spin-spin correlation functions 
in examples~(\ref{eq: example 2}-\ref{eq: example 3}) 
or spatial correlation functions between the local bosonic densities
$n^{\ }_{\langle ij\rangle}$ in example~(\ref{eq: example 4}) 
decay algebraically with separation
in the scaling regime $g/T,T/U\to0$.
This is so because examples~(\ref{eq: example 2}-\ref{eq: example 3}) 
reduce to the non-interacting classical three-coloring model 
that was solved by Baxter in Ref.~\onlinecite{Baxter1970}, 
whereas example~(\ref{eq: example 4}) 
reduces to the non-interacting classical
square lattice dimer model that was solved by Kasteleyn
in Ref.~\onlinecite{Kasteleyn63}. In either case, it is now
understood that the \textit{constrained entropic scaling limit} $g/T,T/U\to0$
is critical in that correlation functions decay as power laws
in space.
%
%

\subsection{\label{subsec: U perturbation to ce critical point}
The scaling limit $g/T\to0$, $0<T/U\ll1$
           }

In this paper, we shall assume that,
if we soften the condition that entropic fluctuations satisfy
the constraint, i.e., if we consider the limit
$g/T\to0$ holding $T/U$ small but finite, we must then impose
a cutoff $\xi^{\ }_{\mathrm{ce}}(T/U)$ 
to the algebraic decay in space of correlation functions.
In effect, we are assuming that the operator 
related to the appearance of defects that violate the constraint imposed 
by the energy scale $U$ is a relevant perturbation 
to the constrained entropic critical point
in the RG sense. 
This assumption is indeed satisfied 
by examples~(\ref{eq: example 2}-\ref{eq: example 4})
but there are no fundamental reasons for 
it to hold for all constrained systems.

If defects cause the system to flow to a generic
unconstrained fixed point with short-range correlations in space, the
correlation length of the system is controlled by the ratio $T/U$ and
can be estimated with the following argument. 
The concentration of
thermally activated defects that violate the constraint is
proportional to $\exp\left(-\alpha^{\ }_{U}U/T\right)$ when $T \ll U$,
where $\alpha^{\ }_{U}>0$ is some non-universal numerical constant
specific to $\hat{H}^{\ }_{U}$,
since it represents the fugacity of defects that enters as
the coupling constant driving the system away from the
\textit{constrained entropic scaling limit}. 
Therefore, the correlation length of the system is
\begin{subequations} 
\label{eq: macro range T for ce power law}
\begin{eqnarray}
\xi^{\ }_{\mathrm{ce}}(T/U)\sim
\mathfrak{a}
\exp
\left(
\frac{\alpha^{\ }_{U}}{d-y^{\ }_{U}}\frac{U}{T}
\right),
\label{eq: xi-ce-(T/U)}
\end{eqnarray} 
when $T \ll U$.
Here, $\mathfrak{a}$ is the lattice spacing,
$d$ is the dimensionality of space, 
and $0<y^{\ }_{U}<d$ is the scaling dimension of the operator
representing a defect. 
Upon approaching the constrained entropic critical point
$T/U=0$, the exponential growth of the correlation length
$\xi^{\ }_{\mathrm{ce}}(T/U)$ guarantees that correlation functions
decay in space as power laws over a macroscopically large window
of length scales 
\begin{eqnarray}
\mathfrak{a}\ll r\ll \xi^{\ }_{\mathrm{ce}}(T/U).
\label{eq: macro range T for ce power law b}
\end{eqnarray}
\end{subequations} 
While the entropic regime~(\ref{eq: regime 3})
of the constrained Hubbard model
is featureless beyond few lattice spacings, 
the entropic regimes~(\ref{eq: regime 3})
of examples~(\ref{eq: example 2}-\ref{eq: example 4}) 
exhibit a scaling behavior over an exponentially large window of
length scales. 
For length scales large compared to the lattice spacing but smaller than
$\xi^{\ }_{\mathrm{ce}}(T/U)$ 
the system behaves as a non-interacting 
classical system with the constraint fully enforced, 
whose correlation functions are captured by a 
(purely entropic) critical theory. 
As the ratio $T/U$ is increased to a value of order 1,
the correlation length
$\xi^{\ }_{\mathrm{ce}}(T/U)$ 
decreases until it becomes of the same order of the lattice spacing.

An interesting possibility occurs if the relevant
coupling constant that drives the system away from an unstable
constrained entropic fixed point generates a 
correlation length 
$\xi^{\ }_{\mathrm{ce}}(T/U)$
that is an algebraic function of
$T/U$ as opposed to an exponential one in $T/U$. If so,
the correlation length 
$\xi^{\ }_{\mathrm{ce}}(T/U)\sim (T/U)^{-\nu^{\ }_{U}}$
for some correlation length exponent $\nu^{\ }_{U}$.
This scaling of the correlation length with temperature 
is identical to that near a quantum critical point, 
$\xi^{\ }_{\mathrm{QCP}}\sim T^{-1/z}$
for some dynamical exponent $z$.
It may thus deceive observers expecting quantum critical scaling: 
the behavior is classical as we are in the temperature regime%
~(\ref{eq: regime 3}), 
very far past the quantum regime~(\ref{eq: regime 1}).
\begin{figure*}
\center
\subfigure{\includegraphics[width=0.95\columnwidth]{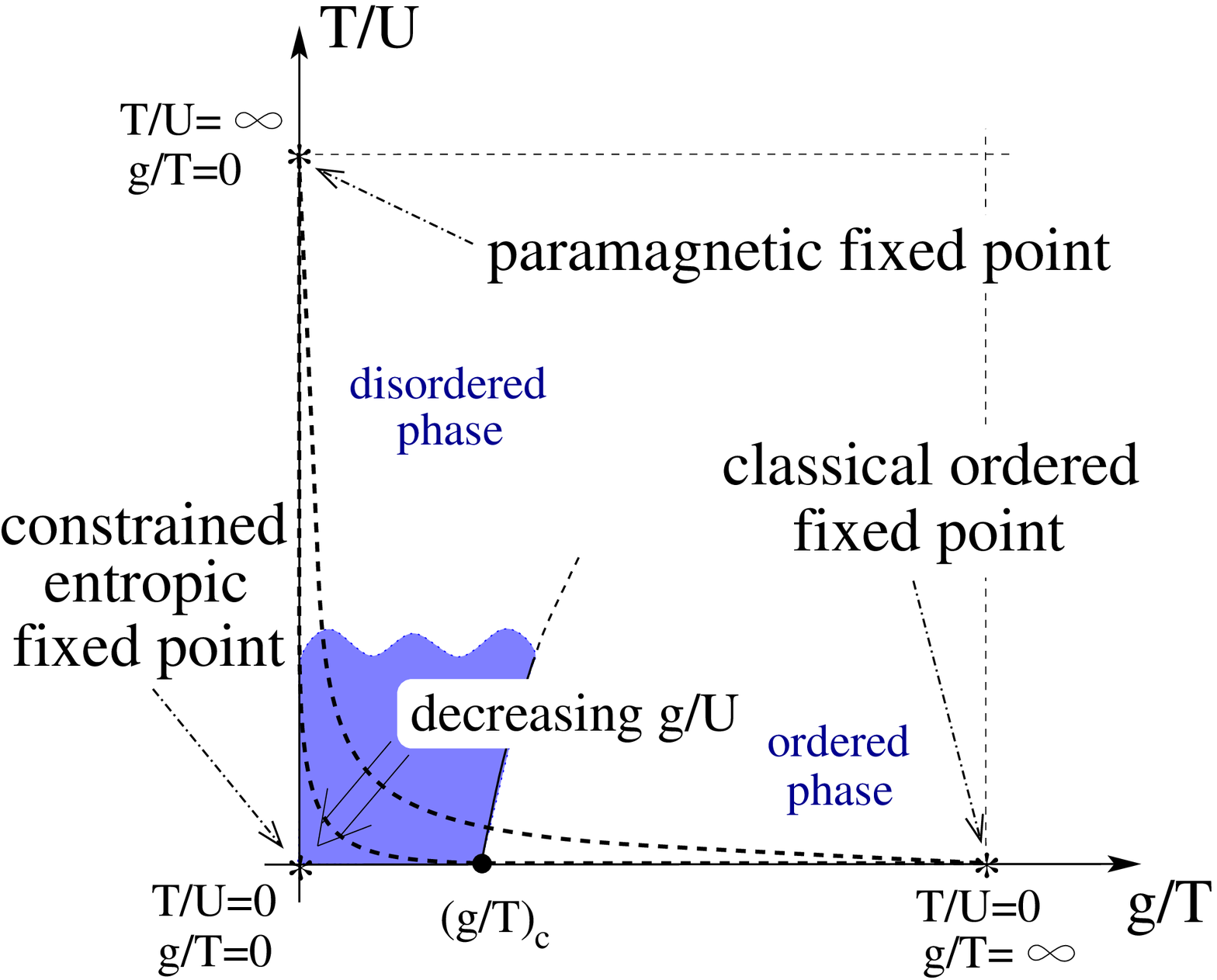}}
\hspace{0.3 cm}
\subfigure{\includegraphics[width=0.95\columnwidth]{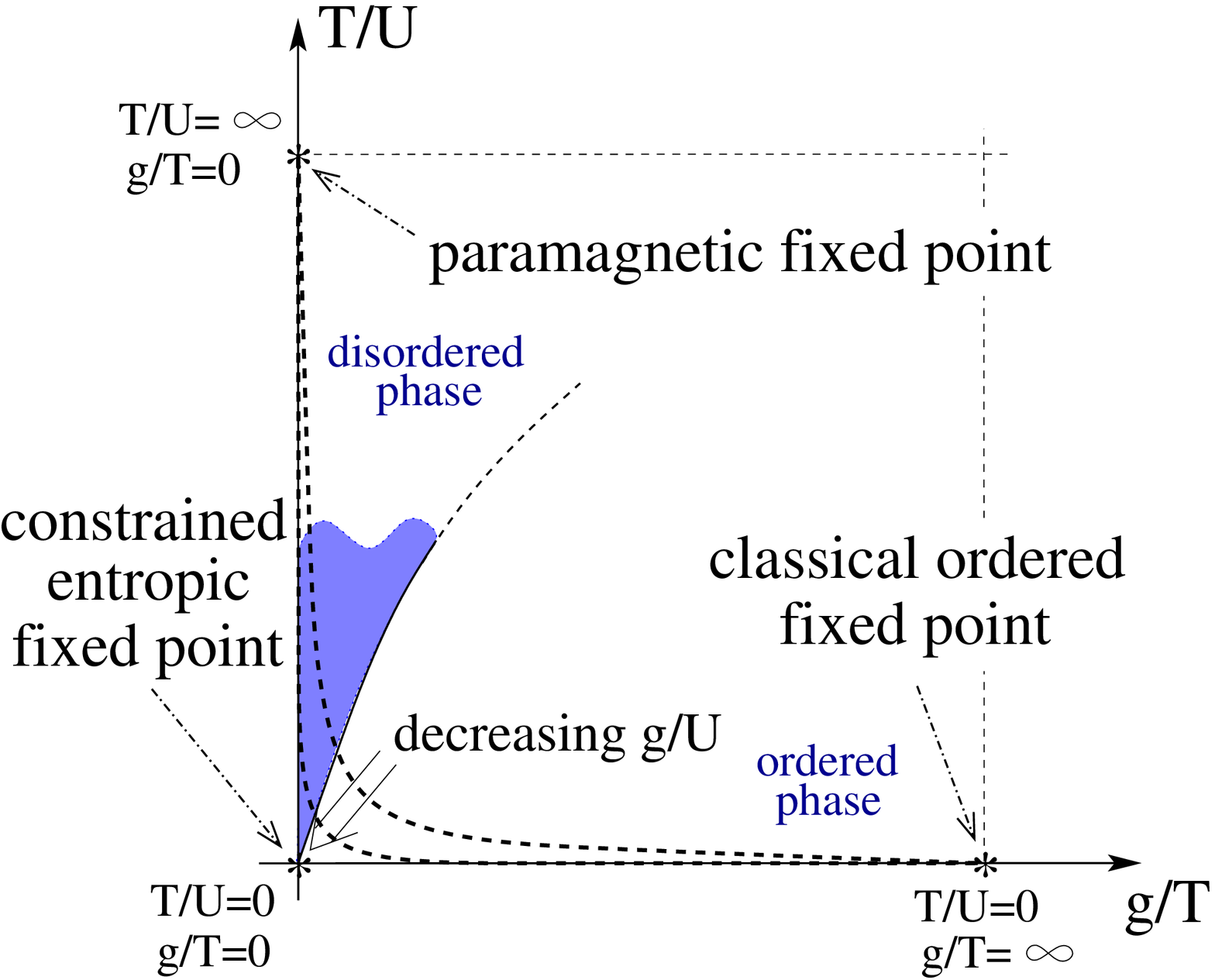}} 
\caption{
\label{fig: cartoon phase diagram} 
(Color Online) -- Qualitative phase diagram of the classical
constrained system assuming the existence of two phases, a classical
ordered phase and a disordered one, separated by a phase boundary. The
details illustrated in these two figures, such as the phase
boundaries, only reflect a close neighborhood of the origin of the
phase diagram, and not features outside this neighborhood.
The dashed lines represent curves where $g$ and $U$ are held fixed and
only the temperature is varied. These curves originate at the ordered
fixed point $(\infty,0)$ at $T=0$ and end at the disordered and
unconstrained entropic fixed point $(0,\infty)$ at $T=\infty$.
In the limit of $g/U \ll 1$, these curves become infinitesimally close
to the $g/T$ and $T/U$ semi-axes.
The universal physics discussed in this paper occurs in
regime~(\ref{eq: regime 3}), therefore in the region $T/U \ll 1$, $g/T
\ll 1$ close to the origin of the coordinate system. The shaded region
in the Figure represents the scaling entropic region appearing in a
constrained system due to the proximity to the constrained entropic
critical point at the origin. (Left) This diagram encompasses
scenarios I, III, and IV depending on the behavior of the system along
the segment $0 \leq g/T \leq (g/T)^{\ }_{\mathrm{c}}$ with $T/U =
0$. (Right) Phase diagram corresponding to scenario II, where
$(g/T)^{\ }_{\mathrm{c}} = 0$.
}
\end{figure*}
%
%

\subsection{\label{subsec: g perturbation to ce critical point}
The scaling limit $T/U\to0$, $0<g/T\ll1$
           }
We now perturb the constrained entropic critical point
$(g/T,T/U)=(0,0)$ by working at a finite value of $g/T$ while
$T/U=0$. We call scenario I the case when $g/T$ is irrelevant if
sufficiently small but becomes relevant beyond some finite critical
value $(g/T)^{\ }_{\mathrm{c}}$.  We call scenario II the case when
any finite value of $g/T$ is relevant.  Scenario III occurs when $g/T$
is exactly marginal up to some critical value $(g/T)^{\
}_{\mathrm{c}}$, in which case the segment $0\leq g/T\leq (g/T)^{\
}_{\mathrm{c}}$ realizes a line of critical points.  At last, scenario
IV happens when the small coupling $g/T$ preserves criticality but
strongly alters its nature; for instance it changes continuously the
value of the central charge in the two-dimensional example given in
Sec.~\ref{sec: constrained Ising J<0}.

In the case of scenario I, all properties of the
constrained entropic critical point
$(g/T,T/U)=(0,0)$ survive a sufficiently small perturbation
$g/T$ until it reaches the critical value
$(g/T)^{\ }_{\mathrm{c}}$
at which a phase transition takes place to a non-critical phase of matter
selected by the characteristic energy scale $g$.
This classical phase transition can be continuous but need not be so.
The non-critical phase of matter could support a conventional classical 
long-range order such as, say, antiferromagnetic order. 
If classical frustration effects are prevalent,
the non-critical phase of matter could support
less conventional classical order such 
as spin-glass order.
An exotic possibility occurs when 
classical frustration effects
select a non-critical phase devoid of any long-range order.

In the case of scenario II, any finite $g/T$ generates a finite
correlation length $\xi^{\ }_{\mathrm{II}}(g/T)$ 
by causing the system to order into a non-critical phase.
This correlation length diverges with $g/T\to0$ in the case of a
continuous classical phase transition at infinite temperature, 
but it may as well remain 
finite, for example if the transition is first order.

Scenarios III and IV differ from scenario I in that
the segment $0\leq g/T \leq(g/T)^{\ }_{\mathrm{c}}$
with $T/U=0$ is a line of critical points in both cases.
The number of critical degrees of freedom  is unchanged
in case III whereas it does change in case IV
as a function of
$0\leq g/T \leq(g/T)^{\ }_{\mathrm{c}}$.

The constrained quantum XXZ Heisenberg and transverse field Ising models%
~(\ref{eq: example 2}) and~(\ref{eq: example 3}), respectively,
provide explicit realizations of scenarios I, II, III, and IV, 
as we will demonstrate in Sec.~\ref{sec: constrained Ising}
and Sec.~\ref{sec: The square lattice dimer model}.
%
%

\subsection{\label{subsec: g and U perturbation to ce critical point}
Perturbing the constrained entropic critical point with $0<T/U$ and $0<g/T$
           }
The fate of the transition at
$\big((g/T)^{\ }_{\mathrm{c}},0\big)$
on the boundary of the phase diagram parametrized by the
dimensionless couplings $(g/T,T/U)$ is model dependent
as one moves to the interior of the phase diagram. 

For simplicity, we assume that the phase diagram consists of two
competing phases only. One phase is the basin of attraction of the
unconstrained entropic stable fixed point located at
$(g/T,T/U)=(0,\infty)$. The other phase is the basin of attraction of
the stable fixed point located at $(g/T,T/U)=(\infty,0)$. A cartoon
version of this phase diagram is depicted in Fig.~\ref{fig: cartoon
phase diagram} (Left) for scenarios I, III, or IV and in
Fig.~\ref{fig: cartoon phase diagram} (Right) for scenario II. Curves
with the dimensionless ratio $g/U$ held fixed are represented by
dashed lines in Fig.~\ref{fig: cartoon phase diagram}.  Changing the
temperature for some given $g/U$ corresponds to moving along a dashed
line in Fig.~\ref{fig: cartoon phase diagram}.  The smaller the ratio
$g/U$, the larger the temperature range for which the system lingers
in the vicinity of the constrained entropic critical point $(0,0)$
along a line with $g/U$ held fixed.

Finally, observe that the location of the phase boundary close to the
constrained entropic critical point $(0,0)$ follows from
\begin{eqnarray}
\xi^{\ }_{\mathrm{II}}(g/T)\sim\xi^{\ }_{\mathrm{ce}}(T/U)
\label{eq: scenario II, T cross}
\end{eqnarray}
in the case of scenario II, assuming that the phase boundary is a line of
continuous transitions. One interesting feature of scenario II is that the
correlation length increases as temperature increases for a large range of
temperatures! The reason is that, since the constrained entropic fixed point
is unstable in both horizontal ($g/T$) and vertical ($T/U$) directions, as
the temperature is raised and one moves in parameter space along the dashed
line with $g/U\ll 1$, the correlation length first increases as one
approaches the fixed point from the horizontal direction, and only starts to
decrease as one moves away along the vertical direction. The crossover
temperature scale is set by Eq.~(\ref{eq: scenario II, T cross}), and it goes
to infinity as $U \to \infty$.
%
%

\section{\label{sec: constrained Ising} 
The constrained quantum XXZ Heisenberg and transverse field Ising models
        }
We are going to illustrate how the high-temperature critical
scaling picture is realized for the constrained quantum XXZ Heisenberg and
transverse field Ising models on the honeycomb lattice. The effective
models are identical for both systems, the only difference lying with the
order $n$ in $\Gamma/U$ needed to generate the effective ``ring
exchange'' $\hat{H}^{(n)}_{\Gamma/U}$ in Eq.~(\ref{eq: H eff}). 
So it suffices to analyze the constrained transverse field
Ising model~(\ref{eq: example 3}) 
whose effective Hamiltonian~(\ref{eq: H eff}) restricted to the subspace
$\mathcal{H}^{\ }_{0,U}$ takes the form
\begin{eqnarray}
\hat{H}^{\ }_{\mathrm{eff}}&:=&
-
h 
\sum_{i=1}^{N} 
\hat{\sigma}^{\rm z}_{i} 
-
h^{\ }_{s} 
\sum_{i=1}^{N} 
(-1)^i
\hat{\sigma}^{\rm z}_{i} 
-
J
\sum_{\langle ij\rangle}
\hat{\sigma}^{\mathrm{z}}_{i}
\hat{\sigma}^{\mathrm{z}}_{j}
\nonumber\\
&&
-
\Gamma_{\textrm{eff}}
\left\{ 
  \sum_{\hexagon}
    \prod_{v=1}^{6} \hat{\sigma}^{\mathrm{x}}_{i^{\ }_{v}}
  +
  \cdots
\right\},
\label{eq: example 3 bis eff}
\end{eqnarray}
where the indices $i^{\ }_{v}, v=1,\dots,6$ label the sites around an
elementary hexagon $\hexagon$. 
The coupling $h$ describes a uniform magnetic field, $h^{\ }_{s}$ describes 
a staggered magnetic field, and $J$ describes an exchange interaction
between nearest-neighbor sites of the honeycomb lattice. 
$J>0$ favors ferromagnetic order while $J<0$ favors antiferromagnetic 
order. Both the classical antiferromagnetic and ferromagnetic ground
states are compatible with the constraint that the magnetization of
each hexagonal plaquette has to be $\pm 6$ or $0$. These couplings,
$h$, $h^{\ }_{s}$, and $J$, are three $g$-like couplings 
that we analyze below.

The temperature in regime~(\ref{eq: regime 3}) is small
compared to $U$ and large compared to 
$|\Gamma^{\ }_{\textrm{eff}}|\propto|\Gamma^{6}/U^{5}|$.  
The approximation of neglecting either
violations of the constraint or the off-diagonal part of the quantum
Hamiltonian should thus be a good starting point. 
If so, in the scaling regime
$|h|/T,|h^{\ }_{s}|/T,|g|/T,T/U\to0$, 
the model reduces to a non-interacting
Ising model on the honeycomb lattice, with the constraint that the
magnetization of each hexagonal plaquette has to be $\pm 6$ or $0$.
This model maps onto the non-interacting three-coloring model 
which was studied by Baxter and whose entropy can be
computed exactly in the thermodynamic limit.~\cite{Baxter1970}  
He also showed that the model exhibits algebraically decaying spatial
correlations and as such is critical. We shall call 
this constrained entropic scaling limit $|h|/T,|h^{\ }_{s}|/T,|g|/T,T/U\to0$
the Baxter critical point. The long-wavelength, low energy limit of
this model is captured by a conformally invariant field theory with
central charge $c=2$.\cite{Kondev96} 

As the temperature is lowered, 
the effects of the couplings $J$, $h$, $h^{\ }_{s}$ 
and any other couplings compatible
with the symmetries of Hamiltonian~(\ref{eq: example 3 bis eff}) need 
to be taken into account.  That is, we need to decide 
what perturbations are relevant, marginal, and irrelevant at the
Baxter critical point. Infinitesimally close to it in parameter space 
one can use perturbative renormalization group arguments. At a
finite distance away 
these methods fail and one must rely on numerical tools to explore the 
stability of the Baxter critical point.

Natural choices for perturbations of the
Baxter critical point are a uniform magnetic field $h$, 
a staggered field $h^{\ }_{s}$, 
and a nearest-neighbor interaction $J$. 
The system is still exactly solvable in the presence of the $h^{\ }_{s}$
coupling alone. In the presence of either the $h$ or $J$
coupling, the system is no longer exactly solvable, and we chose to
resort to numerical transfer matrix calculations in order to
investigate the fate of the Baxter critical point.

Since we are interested in distinguishing between criticality and any
long-range ordered or disordered gapped phase, a convenient choice is
to measure the central charge of the system. This can be obtained from
the coefficient of the largest finite-size scaling correction to the
free energy of a semi-infinite system with periodic boundary
conditions in the finite direction.~\cite{Blote1986,Affleck1986} The
central charge is known to be strictly non-zero if the theory
describing the long-wavelength behavior of the system is critical
(massless). Numerically, a massive phase is signaled by the vanishing
of the measured $c$, for the finite size corrections vanish faster
than in a critical system. (There are conformally invariant
topological field theories with $c=0$, but the numerically measured
$c=0$ is here more trivially an indication of a massive phase.)
{}From similar calculations on the semi-infinite system one can also
obtain the scaling dimensions of the operators in the conformal field
theory describing the long wavelength behavior of the system.  In
addition to providing a better understanding of the critical regime,
the scaling dimensions are known to either vanish or diverge as the
system becomes massive, and they can be used to confirm the central
charge results.

In order to compute the central charge and scaling dimensions of the system 
in a cylindrical geometry, we made use of transfer matrix techniques in 
combination with sparse matrix diagonalization routines from the free 
package ARPACK. 
Our results for the central charge are obtained either as a
function of $h/T$ or as a function of $J/T$. We did not consider the
case of $h$ and $J$ simultaneously present, nor the case when either is
present together with $h^{\ }_{s}$.
%
%

\subsection{Uniform field $h$ -- Scenario I}

\begin{figure}
\center
\includegraphics[width=0.8\columnwidth]{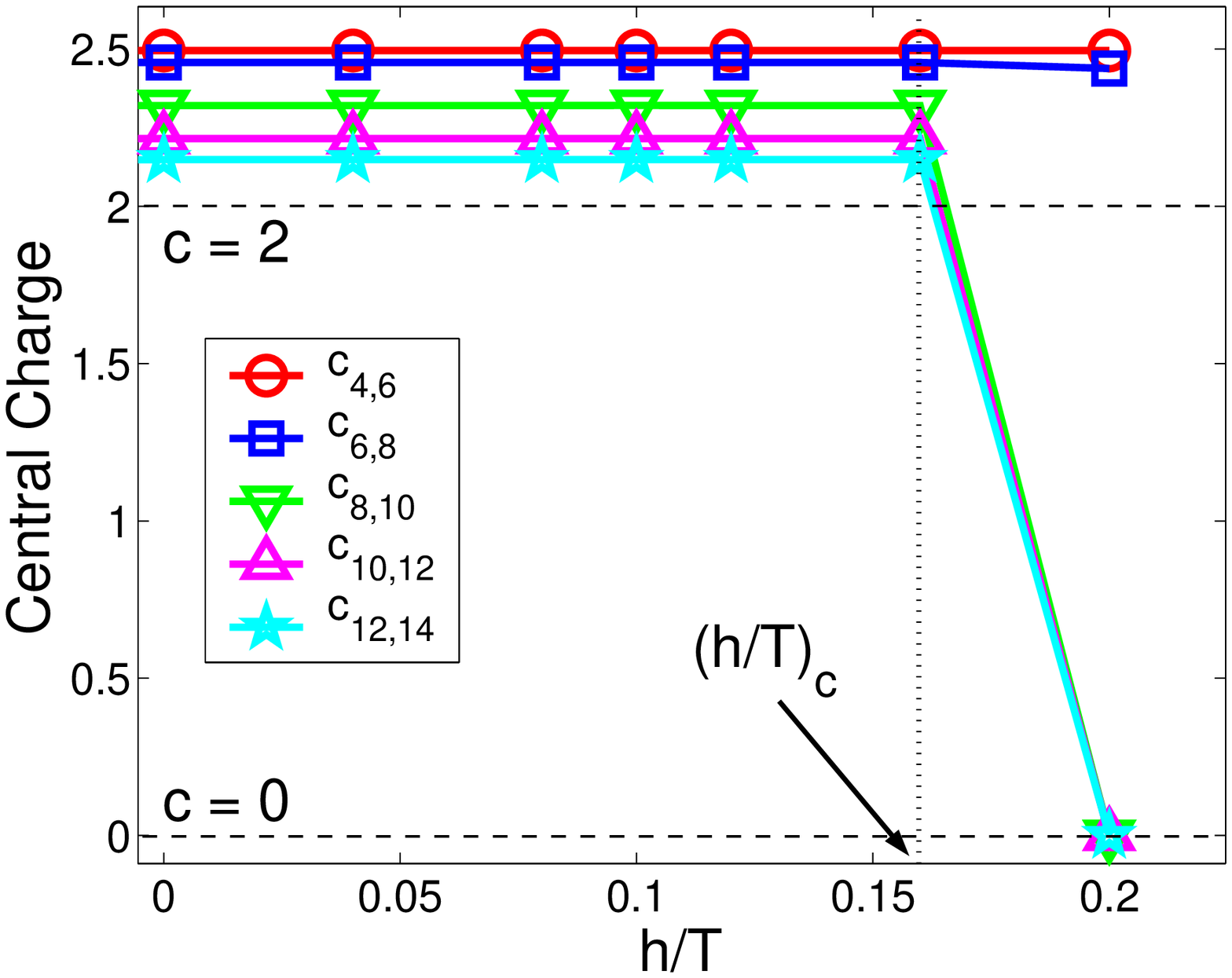}
\includegraphics[width=0.8\columnwidth]{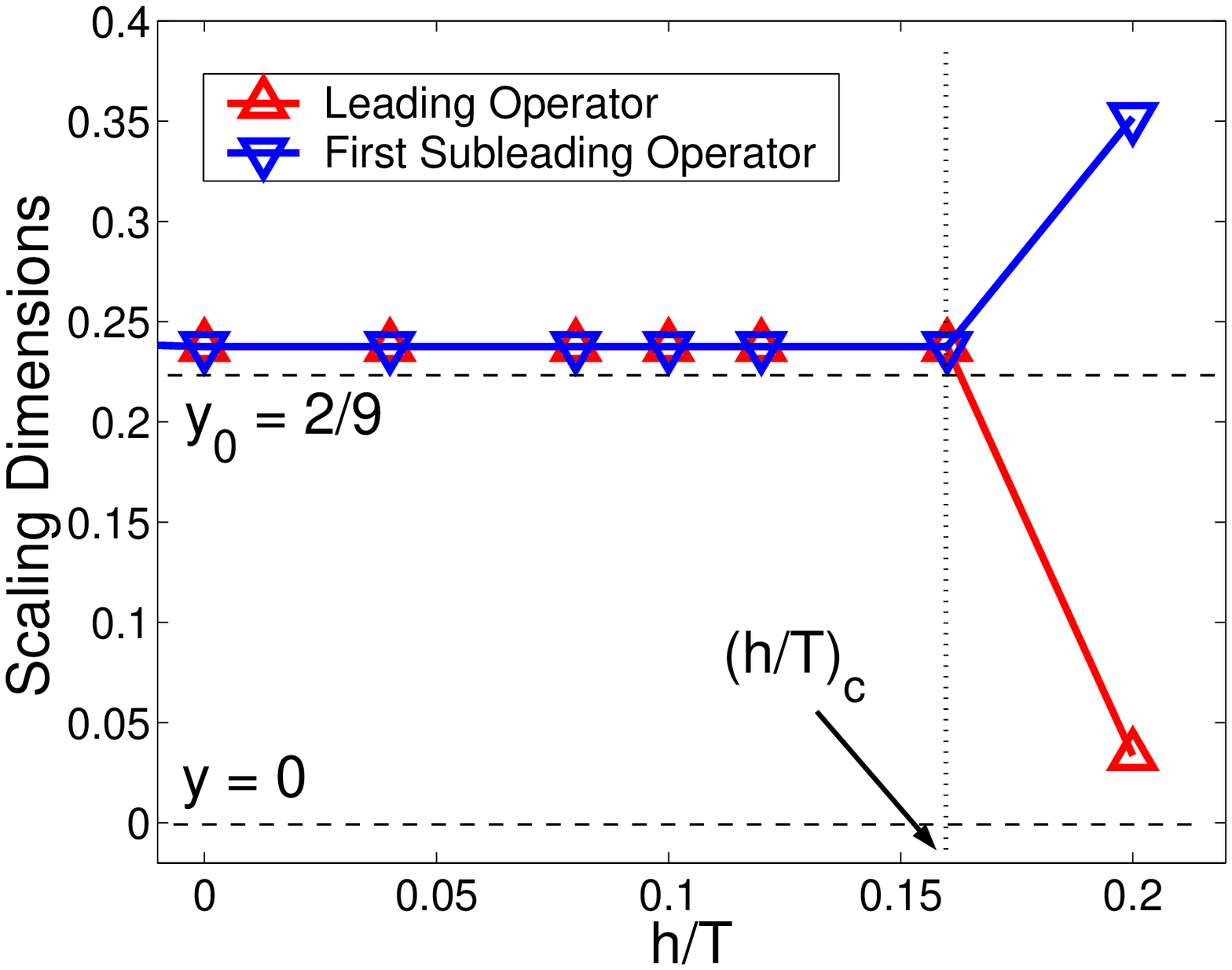}
\caption{
\label{fig: constrained Ising h} 
(Color Online) -- 
(Top) 
Behavior of the (finite size) central charge 
$c^{\ }_{L^{\ }_{i},L^{\ }_{i+1}}$
as a function of $h/T$, obtained 
from finite-size scaling of the free energy computed via transfer matrix. 
{}From top to bottom, the different curves correspond to increasing diameters
$L^{\ }_{j}=4,\cdots,14$ 
for the systems used in the scaling fit 
to compute the value of the central charge $c^{\ }_{L^{\ }_{i},L^{\ }_{i+1}}$. 
The system in presence of a field is symmetric upon the sign change 
$h \leftrightarrow -h$, therefore only the $h>0$ axis is shown. 
Notice that the high-temperature criticality is robust with respect to 
a uniform field and it survives at large but finite temperatures. 
(Bottom) 
Behavior of the two smallest scaling dimensions allowed by the conformal 
field theory as a function of $h/T$. 
The extrapolated values for $L \to \infty$ are shown here for simplicity. 
The fact that both (degenerate) scaling dimensions remain constant in the 
critical regime suggests that the whole critical phase is described by 
the same conformal field theory. 
The lines between data points are guides to the eyes.
        }
\end{figure}

The numerical results for a uniform magnetic field $h$ are presented in 
Fig.~\ref{fig: constrained Ising h}
[as the transformation $h\to-h$ and 
$\hat{\sigma}^{\mathrm{z}}_{i}\to- \hat{\sigma}^{\mathrm{z}}_{i}$
leaves the Hamiltonian unchanged, 
it is sufficient to consider the case $h/T\geq0$].
The dependence on $h/T$ of the central charge of a honeycomb lattice 
wrapped around a cylinder is shown 
in Fig.~\ref{fig: constrained Ising h} (Top), 
for different values of the cylinder radius.
The dependence on $h/T$ of the two smallest scaling dimensions 
is shown in Fig.~\ref{fig: constrained Ising h} (Bottom).

The numerical calculations proceed in two steps, 
to be repeated for each value of the reduced magnetic field $h/T$.%
~\cite{Blote1986,Affleck1986}
First, the three largest eigenvalues 
$\Lambda^{(0)}_{j}\geq\Lambda^{(1)}_{j}\geq\Lambda^{(2)}_{j}$  
of the transfer matrix corresponding to the diameters $L^{\ }_{j} = 2j$, 
$j=2,\dots,7$ of the cylinder are computed. 
Here $L^{\ }_{j}/2$ corresponds to the number of hexagonal plaquettes, or 
equivalently $2 L^{\ }_{j}$ is the number of spins in a row of the infinite 
cylinder. The largest eigenvalue $\Lambda^{(0)}_{j}$
yields the dimensionless free energy per spin 
\begin{subequations}
\begin{eqnarray}
f^{\ }_{j} = - \frac{1}{2 L^{\ }_{j}} \ln(\Lambda^{(0)}_{j}) 
\end{eqnarray}
while the next two subleading eigenvalues  $\Lambda^{(1,2)}_{j}$
yield the dimensionless excitation energies
\begin{eqnarray}
\Delta f^{(k)}_{j} = 
\frac{1}{2 L^{\ }_{j}} 
\ln
\left(
\frac{
\Lambda^{(0)}_{j}
     }
     {
\Lambda^{(k)}_{j}
     }
\right),
\qquad
k=1,2.
\end{eqnarray}
\end{subequations}
Second, a value of the (finite size) central charge 
$c^{\ }_{L^{\ }_{i},L^{\ }_{i+1}}$  
is obtained from the finite-size scaling fit 
performed on two consecutive values of the free 
energy, i.e., on data points of the type 
$\{(L^{\ }_{j},f^{\ }_{j}),\,j=i,i+1\}$. 
This is repeated for $i=2$ to $i=6$. 
The (finite size) scaling dimensions 
$y^{(k)}_{L^{\ }_{i},L^{\ }_{i+1}}$, $k=1,2$, 
follow from finite-size scaling fits 
on two consecutive values of 
$\Delta f^{(k)}_{j}$, 
i.e., on data points of the type 
$\{(L^{\ }_{j},\Delta f^{(k)}_{j}),\,j=i,i+1\}$. 

What we are after is not so much the finite size values
$c^{\ }_{L^{\ }_{i},L^{\ }_{i+1}}$
and
$y^{(k)}_{L^{\ }_{i},L^{\ }_{i+1}}$
as their values in the thermodynamic limit
$L^{\ }_{i} \to \infty$ ($i\to\infty$).
We present finite size data
$c^{\ }_{L^{\ }_{i},L^{\ }_{i+1}}$
in Fig.~\ref{fig: constrained Ising h} (Top)
to illustrate the finite size corrections
while we present the extrapolated values
$y^{(k)}\equiv\lim_{i\to\infty}y^{(k)}_{L^{\ }_{i},L^{\ }_{i+1}}$
in Fig.~\ref{fig: constrained Ising h} (Bottom).
The dependence on $h/T$ of the central charge
$c\equiv\lim_{i\to\infty}c^{\ }_{L^{\ }_{i},L^{\ }_{i+1}}$
that we deduce from 
Fig.~\ref{fig: constrained Ising h} (Top)
is 
\begin{eqnarray}
c=
\left\{
\begin{array}{ll}
2,
&
0\leq
\frac{h}{T}<
\left(\frac{h}{T}\right)^{\ }_{\mathrm{c}},
\\
&
\\
0,
&
\left(\frac{h}{T}\right)^{\ }_{\mathrm{c}}<
\frac{h}{T}.
\end{array}
\right.
\end{eqnarray}

{}From Fig.~\ref{fig: constrained Ising h} one reads 
that the Baxter critical point is robust to the introduction of 
a uniform magnetic field.  
According to Fig.~\ref{fig: constrained Ising h} (Top),
the system remains critical with central charge $c=2$ over the
finite interval $0\leq h/T\leq (h/T)^{\ }_{\mathrm{c}}$, before
entering an ordered phase 
through a (first-order) phase transition at 
$(h/T)^{\ }_{\mathrm{c}}$. 
This is confirmed by the behavior of the scaling dimensions shown in 
Fig.~\ref{fig: constrained Ising h} (Bottom), which seem 
to rapidly vanish/diverge across the transition at 
$(h/T)^{\ }_{\mathrm{c}}$, respectively. 
Also, according to Fig.~\ref{fig: constrained Ising h} (Bottom),
the smallest scaling dimensions are unchanged along the segment
$0\leq h/T\leq (h/T)^{\ }_{\mathrm{c}}$. 
These numerical results support the conclusion that a small coupling 
$h/T$ is irrelevant at the Baxter critical point, and 
scenario I is realized along the segment. 

This behavior is perhaps
surprising if compared to the effect of a uniform magnetic field in
the unconstrained Ising model on the honeycomb lattice. In that case,
the uniform magnetic field is a relevant perturbation that causes the
system to order at any finite temperature. The origin of this
difference is due to the large depletion of configurations with finite
magnetization induced by the projective action of the constraint. The
entropy of the system as a function of magnetization $m$ seems to
acquire a cusp at $m=0$ that leads to a strong first order transition
at finite temperature in the presence of a uniform magnetic field.
%
%

\subsection{
Staggered field $h^{\ }_{s}$ -- Scenario II
           } 
The case of the staggered magnetic field $h^{\ }_{s}$ has been solved
exactly by Baxter~\cite{Baxter1970} in the limit 
$T/U\to0$, $h^{\ }_{s}/T$ arbitrary.  The model exhibits an
infinite order phase transition as $h^{\ }_{s}/T\to0$.  
The staggered field is a marginally relevant coupling 
to the Baxter critical point $h^{\ }_{s}/T,h^{\ }_{s}/U,T/U\to0$,%
~\cite{Kondev96} i.e., a
staggered magnetic field realizes scenario II of 
Sec.~\ref{subsec: g perturbation to ce critical point}.  
In particular, this implies that any small staggered field 
$h^{\ }_{s}/T$ induces an 
exponentially large correlation length
\begin{eqnarray}
\xi^{\ }_{\mathrm{II}}(h^{\ }_{s}/T)\sim
\mathfrak{a}
\exp\left(\alpha^{\ }_{h^{\ }_{s}}\frac{T}{h^{\ }_{s}}\right)
\label{eq: xi-ce-hs/T}
\end{eqnarray} 
where $\alpha^{\ }_{h^{\ }_{s}}$ is some dimensionless number.

The divergence of the  correlation length~(\ref{eq: xi-ce-hs/T})
as $h^{\ }_{s}/T\to0$ is cut off at the crossover temperature 
\begin{eqnarray}
T^{\ }_{\mathrm{cross}}\sim
\sqrt{Uh^{\ }_{s}}. 
\label{eq: crossover T if staggered}
\end{eqnarray}
This estimate follows from the finite correlation length 
induced by the fact that $T/U$, although large, is finite. 
Indeed, the Baxter critical point is
unstable to constraint-violating defects that appear as soon as $T/U$
is finite. The dependence on $T/U$ of the corresponding correlation
length $\xi^{\ }_{\mathrm{ce}}(T/U)$ is governed by the most relevant
operator that introduces violations of the constraint in the spin
language. This operator corresponds to the insertion of fractional
vortices in the continuum theory that describes the Baxter critical
point.  The scaling dimension $y^{\ }_{U}$ of this operator is given
by $y^{\ }_{U}=2/9$ according to Ref.~\onlinecite{Moore2004} in the case
when there are no interactions added to the model. This is indeed in 
agreement with our numerical results for $h/T,J/T\to 0$ 
(see Figs.~\ref{fig: constrained Ising h} and~\ref{fig: constrained
Ising J}). The estimate~(\ref{eq: xi-ce-(T/U)}) thus becomes
\begin{eqnarray}
\xi^{\ }_{\mathrm{ce}}(T/U)&\sim& 
\mathfrak{a}
\exp\left(\alpha^{\ }_{U}\frac{9U}{16T}\right)
\label{eq: xi-ce-T/U if 3 color}
\end{eqnarray}
and the crossover temperature~(\ref{eq: crossover T if staggered})
follows from solving Eq.~(\ref{eq: scenario II, T cross})
with the help of 
Eqs.~(\ref{eq: xi-ce-hs/T}) 
and (\ref{eq: xi-ce-T/U if 3 color}).
%
%

\subsection{
Nearest-neighbor interaction $J>0$ -- Scenario III
           }
\begin{figure}
\center
\includegraphics[width=0.8\columnwidth]{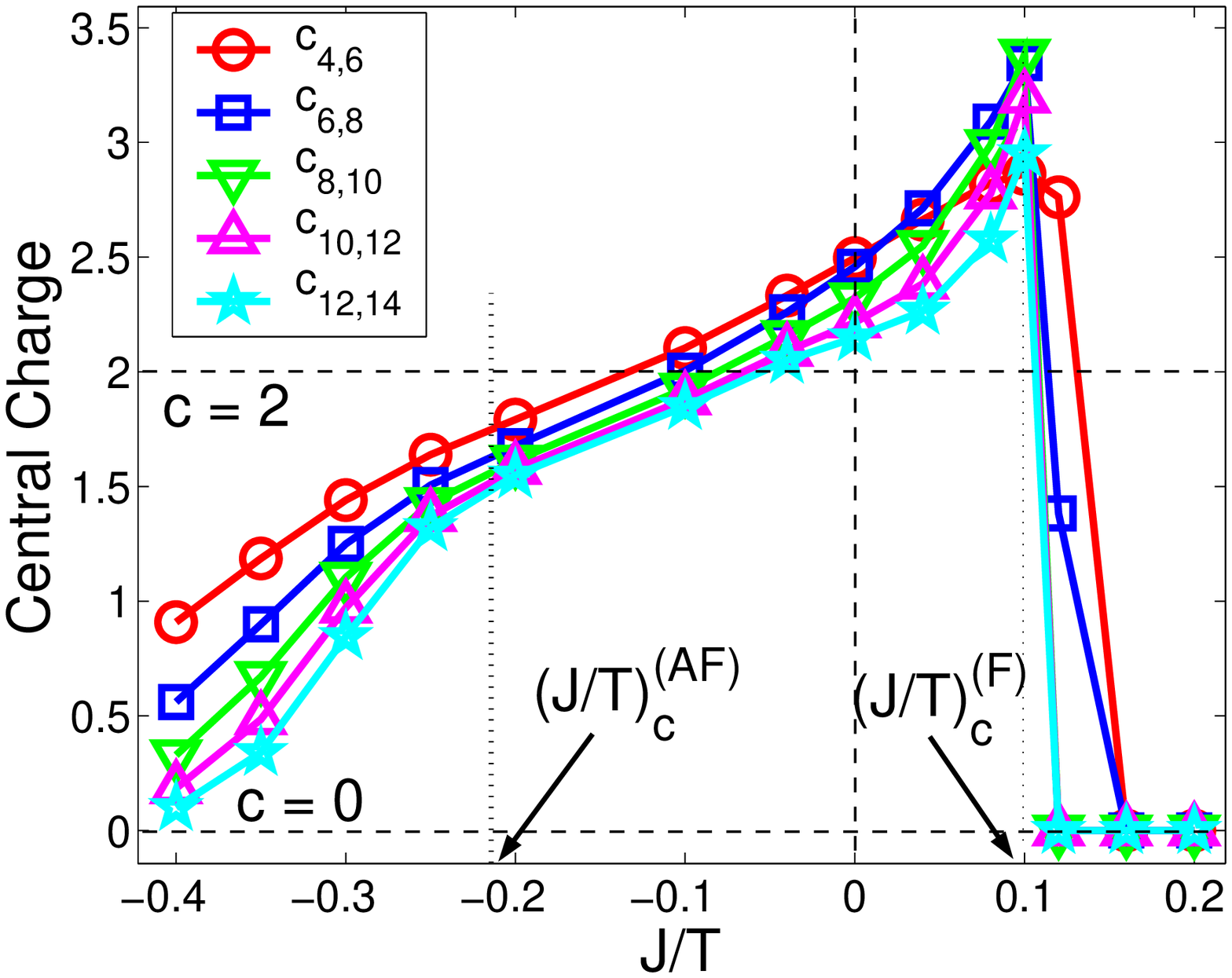}
\includegraphics[width=0.8\columnwidth]{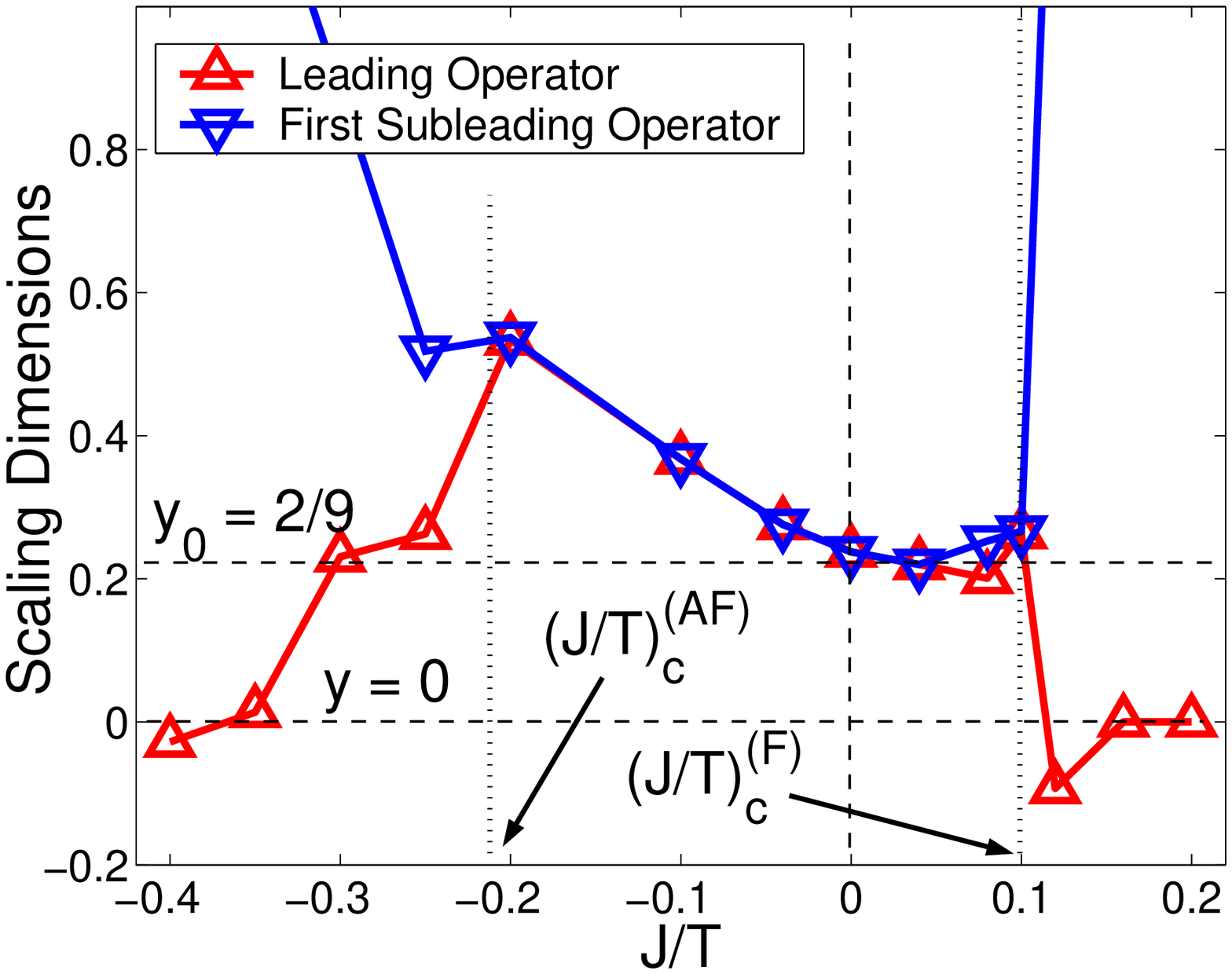}
\caption{
\label{fig: constrained Ising J} 
(Color Online) -- 
(Top) 
Behavior of the (finite size) central charge 
$c^{\ }_{L^{\ }_{i},L^{\ }_{i+1}}$
as a function of $J/T$, obtained 
from finite-size scaling of the free energy computed via transfer matrix. 
{}From top to bottom, the different curves correspond to increasing diameters
$L^{\ }_{j}=4,\cdots,14$ 
for the systems used in the scaling fit 
to compute the value of the central charge $c^{\ }_{L^{\ }_{i},L^{\ }_{i+1}}$. 
Positive values of $J$ correspond to a ferromagnetic coupling, while 
negative values correspond to an antiferromagnetic coupling. 
Notice that the high-temperature criticality is robust with respect to 
uniform nearest-neighbor interactions and it survives at large but finite 
temperatures both for positive and negative $J$. 
Antiferromagnetic interactions however deeply affect the critical behavior, 
inducing what appears to be a continuously varying central charge. 
(Bottom) 
Behavior of the two smallest scaling dimensions allowed by the conformal 
field theory as a function of $J/T$. 
The extrapolated values for $L \to \infty$ are shown here for simplicity. 
Contrary to the uniform field case, the scaling dimensions vary 
on the AF side $(J/T)^{(AF)}_{\mathrm{c}}\leq J/T\leq0$.
Variations
of the scaling dimensions on the F side
$0\leq J/T<(J/T)^{(F)}_{\mathrm{c}}$
cannot be resolved numerically. 
The lines between data points are guides to the eyes.
        }
\end{figure}

The numerical results for a uniform nearest-neighbor interaction $J$ 
between the Ising spins are presented
in Fig.~\ref{fig: constrained Ising J}. 
The dependence on $J/T$ of the central charge of a honeycomb lattice wrapped 
around a cylinder is shown in
Fig.~\ref{fig: constrained Ising J} (Top), 
for different values of the cylinder radius.
The dependence on $J/T$ of the two smallest scaling dimensions 
is shown in Fig.~\ref{fig: constrained Ising J} (Bottom).

We consider first the ferromagnetic side of the interaction, $J>0$.
According to Fig.~\ref{fig: constrained Ising J} (Top),
for small but finite $J/T$, the central charge remains larger than or
equal to the noninteracting value $c=2$,
until the central charge drops to zero abruptly.  
This suggests that the correlation length
remains infinite over the finite interval 
$0\leq J/T<(J/T)^{(F)}_{\mathrm{c}}$, until the system undergoes a strong 
first order transition at $(J/T)^{(F)}_{\mathrm{c}}$. 
This is confirmed by the behavior of the scaling dimensions, which seem 
to rapidly vanish/diverge across the transition at 
$(J/T)^{(F)}_{\mathrm{c}}$, respectively. 
Moreover, Fig.~\ref{fig: constrained Ising J} (Bottom) suggests that 
the lowest scaling dimensions vary continuously with $J/T$, even 
though numerics alone cannot be deemed conclusive on this issue. 

The behavior of the central charge in the critical range 
$0\leq J/T<(J/T)^{(F)}_{\mathrm{c}}$ is different from the uniform field 
case, as $c^{\ }_{L^{\ }_{i},L^{\ }_{i+1}}$ is seen to grow significantly upon
approaching the critical value $(J/T)^{(F)}_{\mathrm{c}}$ for a fixed
system size $L^{\ }_{i}$. 
This growth can be explained by exponential corrections
to finite size scaling that are known to occur at a first-order phase
transition. 
Indeed, extrapolating the curves in Fig.~\ref{fig: constrained Ising J} 
in the limit $L^{\ }_{j}\to\infty$ yields an approximately constant value 
of $c=2$ over the interval $0\leq J/T<(J/T)^{(F)}_{\mathrm{c}}$. 
The extrapolated curve is however rather noisy due to the limited range 
of numerically accessible system sizes and it is not shown here. 
The reason why these corrections are so strong in the
case of a nearest-neighbor ferromagnetic perturbation compared to the
case of a uniform magnetic field deserves further study.

Our interpretation of Fig.~\ref{fig: constrained Ising J}
is that, in the thermodynamic limit, 
the central charge is constant on the segment
$0\leq J/T<(J/T)^{(F)}_{\mathrm{c}}$
while the scaling dimensions are not.
If so, the segment
$0\leq J/T<(J/T)^{(F)}_{\mathrm{c}}$
realizes a line of critical points, i.e., 
$J/T$ is a marginal interaction along this segment, 
thus realizing scenario III. This interpretation agrees
with the perturbative RG calculation from Ref.~\onlinecite{Castelnovo2004}. 

The equilibration properties of the constrained XXZ Heisenberg or 
transverse field Ising model in the ferromagnetically ordered phase exhibit 
rather peculiar features, and have been discussed 
in Ref.~\onlinecite{Castelnovo2005}. 
Based on those results, we expect quantum glassiness to appear in the 
system when the temperature is lowered across the transition to the F 
ordered phase, at least as long as the transition temperature is small enough 
for the $U$-violating defects not to play a significant role in the 
equilibration process (Fig.~\ref{fig: simple case}). 
%
%

\subsection{\label{sec: constrained Ising J<0}
Nearest-neighbor interaction $J<0$ -- Scenario IV
           }
The behavior of the central charge is even more surprising on the
antiferromagnetic side of the interaction $J<0$.  The values  
taken by $c^{\ }_{L^{\ }_{i},L^{\ }_{i+1}}$ in
Fig.~\ref{fig: constrained Ising J} (Top) are seen to drop below $c=2$ for
$J/T<-0.1$ as soon as the system size is sufficiently large.
As the diameter of the cylinder is increased 
from $L^{\ }_{i}=4$ to $L^{\ }_{i}=14$, the dependence
of $c^{\ }_{L^{\ }_{i},L^{\ }_{i+1}}$ on $J/T$
seems to indicate the existence of two distinct regimes for $J/T$. 
On the one hand, the $c^{\ }_{L^{\ }_{i},L^{\ }_{i+1}}$ appear to collapse 
onto a nonvanishing value of $c$ for sufficiently large $J/T$ 
(close to $J/T=0$) in the thermodynamic limit. 
On the other hand, consecutive $c^{\ }_{L^{\ }_{i},L^{\ }_{i+1}}$ remain well
separated from each other for sufficiently small (negative) $J/T$, thereby
suggesting a vanishing limiting value $c$ for the central charge 
in the thermodynamic limit. 

We interpret our finite size simulations as signaling the existence of
a (continuous) transition at a finite $(J/T)^{(AF)}_{\mathrm{c}}\simeq
-0.2$, which is in agreement with variational mean-field results by
Cirillo~\textit{et al} in Ref.~\onlinecite{Cirillo1996}. Our
simulations for the two smallest scaling dimensions also agree with
this interpretation. While they are degenerate and they vary
continuously in the temperature region
$(J/T)^{AF}_{\mathrm{c}}<J/T<0$, they split and rapidly vanish/diverge
as soon as $J/T<(J/T)^{(AF)}_{\mathrm{c}}$, respectively. As for our
numerical results for the central charge, we interpret them as
indicative of one of three different possibilities. The first
possibility is that the phase transition at
$(J/T)^{(AF)}_{\mathrm{c}}$ separates a phase with vanishing central
charge below $(J/T)^{AF}_{\mathrm{c}}$ and a line of critical points
with continuously varying central charge above
$(J/T)^{(AF)}_{\mathrm{c}}$ that interpolate between the values
$c=3/2$ at $(J/T)^{(AF)}_{\mathrm{c}}$ and $c=2$ at infinite
temperature. The second possibility is that the finite size data when
$(J/T)^{(AF)}_{\mathrm{c}}<J/T<0$ are the signature of an infinite
sequence of stepwise increases of the central charge in the
thermodynamic limit that interpolates between the values $c=3/2$ at
$(J/T)^{(AF)}_{\mathrm{c}}$ and $c=2$ at infinite temperature.  (See
Ref.~\onlinecite{Fradkin2004} for a possibly related phenomenon.)  And
finally, the third possibility is that the finite size data for $c^{\
}_{L^{\ }_{i},L^{\ }_{i+1}}$ collapse in the thermodynamic limit to
the central charge $c=0$ when
$J/T<\left(J/T\right)^{(AF)}_{\mathrm{c}}$ and $c=3/2$ when
$\left(J/T\right)^{(AF)}_{\mathrm{c}}<J/T<0$.  The plateau with
$c=3/2$ would correspond to a conformal field theory built out of
three Majorana fermions and endowed with a supersymmetry. More
accurate simulations, i.e., simulations that can access larger
diameters of the cylinder on which the honeycomb lattice is wrapped,
are required to select which of these possibilities corresponds to
the correct thermodynamic limit.

\begin{figure}[ht]
\center
\includegraphics[width=.95\columnwidth]{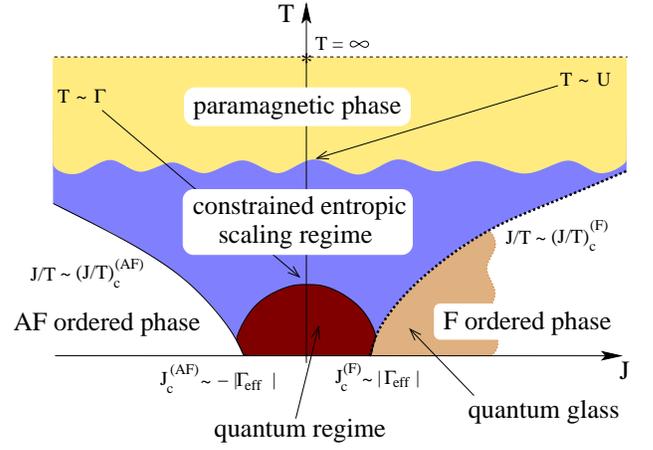}
\caption{
\label{fig: simple case} 
(Color Online) -- 
Illustration of the phase diagram of the constrained Ising model in the 
presence of a nearest-neighbor coupling $J$, discussed in 
Sec.~\ref{sec: constrained Ising}. 
$U$ is held fixed in the figure, and only two out of the three parameters 
of the model, namely $J$ (horizontal axis) and $T$ (vertical axis), 
are shown. 
The effects of the quantum energy scale $\Gamma$ become dominant only at 
very small values of the coupling $J$, in the interval 
$-|\Gamma_{\textrm{eff}}| \lesssim J \lesssim |\Gamma_{\textrm{eff}}|$, 
and for very small 
temperatures $T \lesssim |\Gamma_{\textrm{eff}}|$, giving rise to the 
quantum regime shown in the picture. 
At low temperatures, but for $|J| \gg |\Gamma_{\textrm{eff}}|$ the system 
exhibits two classical ordered phases, a ferromagnetic (F) one for $J>0$ 
and an antiferromagnetic (AF) one for $J<0$. 
While the transition to the AF ordered phase is continuous, 
the transition to the F ordered phase is strongly first order (dotted line). 
As discussed in Ref.~\onlinecite{Castelnovo2005}, quantum glassiness is 
expected to appear in the system when the temperature is lowered across 
the transition to the F ordered phase, at least as long as the transition 
temperature is small enough for the $U$-violating defects not to play a 
significant role in the equilibration process. 
         }
\end{figure}

The phase diagram for the constrained Ising model in the presence 
of a nearest-neighbor interaction is summarized in 
Fig.~\ref{fig: simple case}. 
%
%

\subsection{
Possible experimental realization of the constrained Ising model
in a transverse field
           }
A possible physical realization of the constrained classical Ising model
has been discussed in Refs.~\onlinecite{Castelnovo2004,Moore2004}
in the form of lattices of superconducting devices 
with broken time-reversal symmetry. 
Finding an experimental probe mimicking a staggered magnetic 
field might allow the observation of 
a correlation length that increases with temperature
over a large window of temperatures in regime~(\ref{eq: regime 3}).
%
%

\section{\label{sec: The square lattice dimer model}
The square lattice dimer model
        } 
We are now going to illustrate how the high-temperature critical 
scaling picture is realized for the bosonic model in 
example~(\ref{eq: example 4}). 
The effective Hamiltonian~(\ref{eq: H eff})
restricted to the subspace $\mathcal{H}^{\ }_{0,U}$ 
takes the form~\cite{Fradkin1991} 
\begin{eqnarray}
\hat{H}^{\ }_{\mathrm{eff}}&:=&
v
\sum_{i=1}^{N}
\left(
\hat{n}^{\ }_{i,i+\boldsymbol{x}}
\,
\hat{n}^{\ }_{i+\boldsymbol{y},i+\boldsymbol{y}+\boldsymbol{x}}
+
\boldsymbol{x}
\leftrightarrow
\boldsymbol{y}
\right)
\nonumber\\
&&
-
\Gamma_{\textrm{eff}}
\sum_{i=1}^{N}
\left[
e^{
+\mathrm{i}
\left(
\hat{a}^{\ }_{i,i+\boldsymbol{x}}
+
\hat{a}^{\ }_{i+\boldsymbol{y},i+\boldsymbol{y}+\boldsymbol{x}}
\right)
-\mathrm{i}
\left(
\boldsymbol{x}
\leftrightarrow
\boldsymbol{y}
\right)
  }
+
\mathrm{H.c.}
\right]
\nonumber\\
&&
+
\dots 
\label{eq: example 4 bis eff}
\end{eqnarray}
Here, $\hat{a}^{\ }_{i,i+\boldsymbol{e}}$ is the Hermitian operator 
canonically conjugate to the local bosonic number operator
$
\hat{n}^{\   }_{i,i+\boldsymbol{e}}
$, i.e.,
their commutator is the $\mathbb{C}$-number $\mathrm{i}$. 
Under the assumption that $|v| \sim |t| \ll U$ the coupling 
$\Gamma_{\textrm{eff}} \propto t^2/U$ of the off-diagonal term in the 
effective Hamiltonian~(\ref{eq: example 4 bis eff}) is much smaller 
than $v$. If we relax this assumption by allowing for all possible values of 
the ratio $v/\Gamma_{\textrm{eff}}$,
Eq.~(\ref{eq: example 4 bis eff}) 
is nothing but the square-lattice quantum dimer model introduced by
Rokhsar and Kivelson in Ref.~\onlinecite{Rokhsar1988}.

We are after the high-temperature universal behavior of
regime~(\ref{eq: regime 3}). To this purpose, we consider first the
constrained entropic scaling limit $|v|/T,T/U\to0$ for which
the model reduces to the square-lattice classical non-interacting dimer
model, which was studied 
by Kasteleyn in Ref.~\onlinecite{Kasteleyn63}. 
He showed that the entropy can be computed exactly in the thermodynamic limit.
He also showed that the model exhibits algebraically decaying spatial 
correlations, and, as such, it is critical. We shall call the 
constrained entropic scaling limit $|v|/T,T/U\to0$ 
the Kasteleyn critical point. 

The Kasteleyn critical point is captured by a height model 
which, in the long wave-length limit, is described by the
two-dimensional conformally invariant field theory~\cite{Fradkin1991} 
\begin{subequations}
\label{eq: c=1 action}
\begin{eqnarray}
S= 
\pi K \int d^2 \boldsymbol{x}~\vert\boldsymbol{\nabla}\phi\vert^2 
\label{eq: c=1 action a}
\end{eqnarray}
with stiffness specified by
\begin{eqnarray}
K=\frac{1}{2}
\label{eq: c=1 action b}
\end{eqnarray}
and central charge 
\begin{eqnarray}
c=1.
\label{eq: c=1 action c}
\end{eqnarray}
\end{subequations}

A microscopic dimer is represented in the field theory%
~(\ref{eq: c=1 action})
by a linear combination of two field operators. The first one
is the charge $q^{\ }_{e}=\pm1$ ``vertex operator''
$\exp\big(\mathrm{i}2\pi q^{\ }_{e}\phi\big)$
where the sign assignment has to do with
the lattice being bipartite.
The second one is the ``dipolar operator''
$\boldsymbol{e}\cdot\boldsymbol{\nabla}\phi$
where $\boldsymbol{e}$ is one of the two basis vectors of the square 
lattice such that $\big( i, i+\boldsymbol{e} \big)$ is the pair of sites
covered by the dimer (for some site $i$). 
The two-point correlation function 
\begin{eqnarray}
\left\langle
e^{
\mathrm{i}2\pi
\big(
\phi(\boldsymbol{x})
-
\phi(\boldsymbol{y})
\big)
  }
\right\rangle^{\ }_{K}
\sim
\left(\frac{\mathfrak{a}}{|\boldsymbol{x}-\boldsymbol{y}|}\right)^{1/K}
\label{eq: 2-point magnetic vertex correlator}
\end{eqnarray}
decays with the exponent $2$ when $K=1/2$. 
The two-point correlation function
\begin{eqnarray}
\left\langle
  \left(\vphantom{e^{\big(0\big)}}
    \boldsymbol{e}\cdot\boldsymbol{\nabla}\phi(\boldsymbol{x}) 
  \right)
  \left(\vphantom{e^{\big(0\big)}} 
    \boldsymbol{e}\cdot\boldsymbol{\nabla}\phi(\boldsymbol{y})
  \right) 
\right\rangle^{\ }_{K}
\sim
\left(\frac{\mathfrak{a}}{|\boldsymbol{x}-\boldsymbol{y}|}\right)^{2}
\end{eqnarray}
decays with the exponent $2$ for any stiffness $K>0$.
The value~(\ref{eq: c=1 action b})
is thus special in that the scaling dimensions of the
charge $q^{\ }_{e}=\pm1$ vertex and dipolar operators
are degenerate and equal to 1.

A monomer at site $i$ of the square lattice is a defect in the dimer
covering since site $i$ is not the end point of a dimer.
A finite concentration of monomers is represented in the field theory
~(\ref{eq: c=1 action}) by the local charge $q^{\ }_{m}=\pm 1$ vertex
operator $\exp\big(\mathrm{i}2\pi K q^{\ }_{m}\varphi\big)$ for the
field $\varphi$ dual to $\phi$,
\begin{eqnarray}
\left(
\begin{array}{cc}
\frac{\partial\varphi}{\partial x},
&
\frac{\partial\varphi}{\partial y}
\end{array}
\right)=
\left(
\begin{array}{cc}
\frac{\partial\phi}{\partial y},
&
-
\frac{\partial\phi}{\partial x}
\end{array}
\right).
\end{eqnarray}
The two-point correlation function
\begin{eqnarray}
\left\langle
e^{
\mathrm{i}2\pi K
\big(
\varphi(\boldsymbol{x})
-
\varphi(\boldsymbol{y})
\big)
  }
\right\rangle^{\ }_{K}
\sim
\left(\frac{\mathfrak{a}}{|\boldsymbol{x}-\boldsymbol{y}|}\right)^{K}
\end{eqnarray}
decays with the exponent $1/2$ when $K=1/2$. 
The charge $\pm1$ monomer is represented by
a strongly relevant operator with scaling dimension $1/4$.

In the following we will focus on the temperature region where the
quantum effects are negligible for the thermodynamic properties of the
system, i.e., when $|\Gamma^{\ }_{\textrm{eff}}| \ll T \ll U$.
%
%

\subsection{
The case $v=0$ in bipartite lattices -- Scenarios II and III
           }
It is useful to first consider the situation when $v=0$, and add some
other perturbations to dimer models like those recently studied by
Sandvik and Moessner in Ref.~\onlinecite{Sandvik05}. Introduce a small
fugacity for dimers to cover bonds that, although of finite length,
extend beyond nearest-neighbor bonds. Consider separately two cases:
1) when the long bonds break the bipartiteness of the lattice by
connecting sites in the same sublattices, and 2) when the bonds
preserve the bipartite nature of the lattice.

\begin{figure}
\center
\includegraphics[width=0.8\columnwidth]{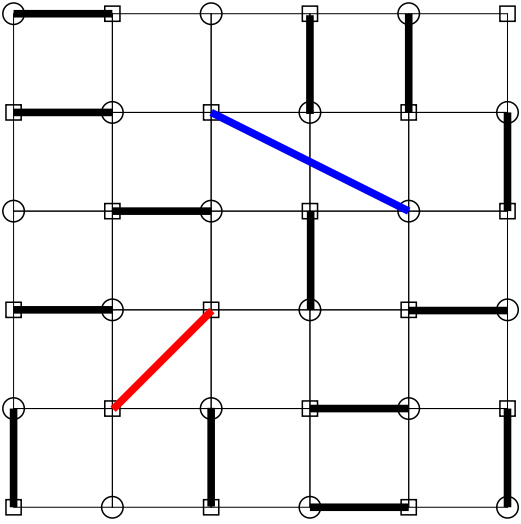}
\caption{
\label{fig:chained-monomers}
(Color Online) -- 
Example of longer dimers that can be thought of as 
pairs of monomers chained together. When the endpoints of a 
long dimer sit at sites within the \textit{same sublattice}, 
the chained monomers have \textit{equal charges}. When the endpoints sit 
at \textit{opposite sublattices}, 
the chained monomers have 
\textit{opposite charges}. 
}
\end{figure}

We interpret these perturbations within the framework of the field theory
as follows. The introduction of appropriate chemical potentials 
leads to a small fraction of the longer dimers. If one of these longer
dimers is simply removed from a configuration, 
one is left with two monomers in the system. 
So a longer dimer can be thought of as two monomers at its
endpoints that are chained together (see
Fig.~\ref{fig:chained-monomers}).
In case 1) a longer dimer connecting sites in the \textit{same
sublattice} can be interpreted as a pair of chained monomers with {\it
equal charges}. The operator product expansion (OPE) of the
corresponding charge $q^{\ }_{m}=\pm 1$ vertex operators leads to
charge $q^{\ }_{m}=\pm 2$ vertex operators. These composite operators
have scaling dimension $2^2\times K/2=1$ when $K=1/2$. Thus, they are 
relevant and they open a gap in the system. The underlying structure
consists of a disordered arrangement of dimers with short range
correlations.  Hence, the Kasteleyn critical point should be strongly
unstable to perturbations that break the sublattice
symmetry.~\cite{Moessner02} The simplest perturbation of this type is
to allow next-nearest neighbor bonds to be covered by dimers. Other
analytical arguments and numerics are consistent with this
expectation.~\cite{Sandvik05,Moessner02}. Therefore, scenario II is
realized upon the introduction of any finite chemical potential for
second-neighbor dimers in the system.

In case 2) a longer dimer connecting sites in \textit{opposite
sublattices} can be interpreted as a pair of chained monomers with
\textit{opposite charges}. The OPE of the corresponding vertex operators
(when all the possible directions of the long dimer are added) will
lead to a $\vert\boldsymbol{\nabla}\phi\vert^2$ term that renormalizes
the stiffness. This is also consistent with the numerical results of
Refs.~\onlinecite{Sandvik05,Moessner02}. 
Scenario III is thus realized. 

We would like to note that violations of the constraint in the form of
monomers are relevant perturbations that drive the system away from
the the Kasteleyn critical point, as in all the situations discussed
is Sec.~\ref{sec: Constrained entropic scaling regime}. The finite
correlation length~(\ref{eq: xi-ce-(T/U)}) thereby generated is given
by
\begin{eqnarray}
\xi^{\ }_{\mathrm{ce}}(T/U)\sim
\mathfrak{a}
\exp
\left(
\alpha^{\ }_{U}\frac{4U}{7T}
\right)
\label{eq: xi-ce-(T/U) square lattice dimer}
\end{eqnarray} 
when $T\ll U$. 
%
%

\subsection{
The case $v \neq 0$ in bipartite lattices -- Scenario III
           }
We are going to argue on the basis of numerics that scenario III can
also be realized by perturbing the Kasteleyn critical point with the
interaction $v$ in Eq.~(\ref{eq: example 4 bis
eff}).

To this end, observe that $\hat{H}^{\ }_{v}$, 
defined by Eq.~line~(\ref{eq: example 4 a}),
counts the total number of elementary square plaquettes of the square
lattice that are flippable in the dimer basis representation.
Here, a flippable plaquette is an elementary square plaquette that has two 
occupied edges (dimers). 
This model has already been studied by Alet~\textit{et al.} in Ref.%
~\onlinecite{Alet05} 
for negative values of the coupling constant $v$. Here we consider both 
positive and negative values, and good agreement with the previous results 
is found where they overlap.\cite{Alet05} 

As exact analytical results are no longer available, we use a numerical 
approach similar to the one described in Sec.~\ref{sec: constrained Ising}. 
We compute the central charge of the system as well as the scaling 
dimensions of two specific operators via transfer matrix techniques, 
using finite size scaling fits. The accessible system sizes are
$L^{\ }_{j} = 2j$, $j=3,\dots,8$ where $L^{\ }_{j}$ is the 
number of square plaquettes across the periodic direction of the system, 
or equivalently $2 L^{\ }_{j}$ is the number of edges. 
As discussed by Alet \textit{et al} in Ref.~\onlinecite{Alet05}, 
the critical regime of this system is captured by a 
$c=1$ two-dimensional conformal field theory of the Coulomb gas type,
with continuously varying stiffness. 
They also computed the scaling dimensions of 
some known operators in the conformal field theory, namely those 
corresponding to the electric and magnetic vortices in the Coulomb gas 
picture. From these measurements one can then obtain the values of all other 
scaling dimensions present in the conformal field theory and discuss the 
nature of the critical phase and of the phase transitions. 

\begin{figure}
\center
\includegraphics[width=0.8\columnwidth]{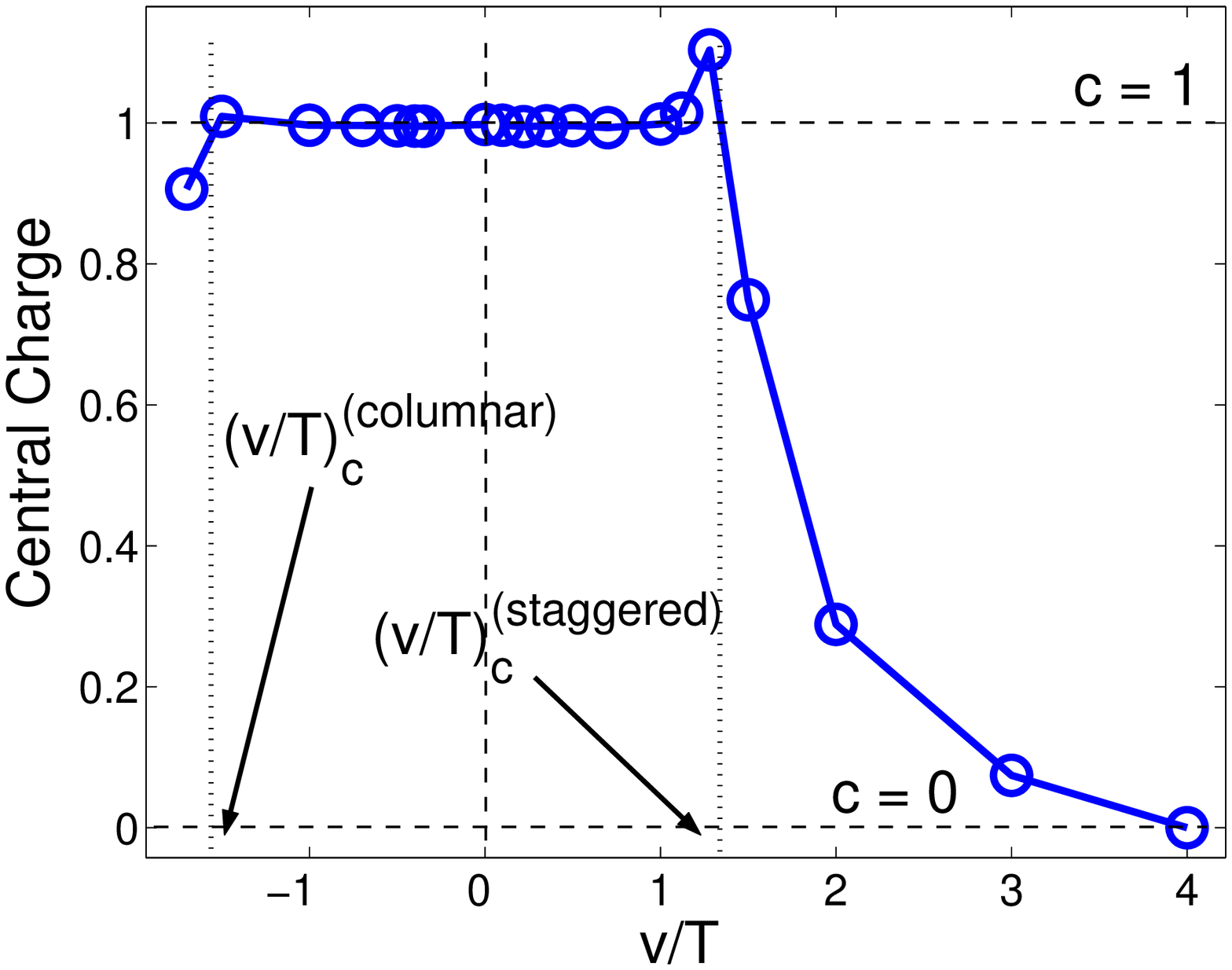}
\includegraphics[width=0.8\columnwidth]{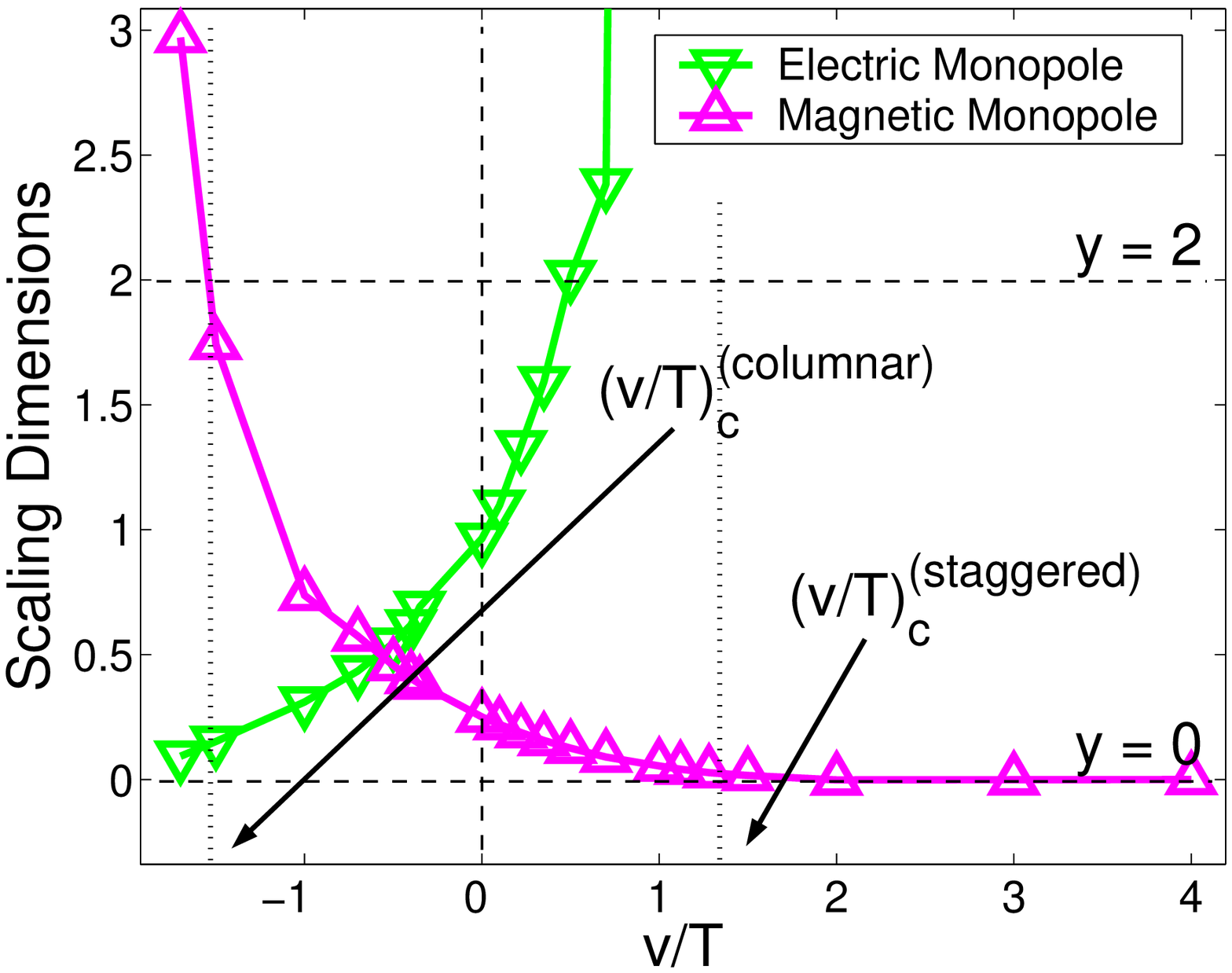}
\caption{
\label{fig: dimer Nc} 
(Color Online) -- 
(Top) 
Behavior of the extrapolated central charge as a function of $v/T$, 
obtained from finite-size scaling of the free energy computed via 
transfer matrix. 
(Bottom) 
Behavior of the scaling dimensions corresponding to the electric and 
magnetic vertex operators as a function of $v/T$, also obtained 
from finite-size scaling arguments and transfer matrix calculations. 
Observe that the scaling dimensions of the electric and magnetic
vertex operators are 1 and 1/4, respectively, when $v/T=0$.
The lines between data points are guides to the eyes.
        }
\end{figure}

The central charge as a function of the coupling constant $v/T$
is plotted in Fig.~\ref{fig: dimer Nc} (Top). 
Thanks to the good convergence in the $L\to\infty$ limit, we report 
the extrapolated values of $c$ instead of each separate curve for 
increasing system size. 
{}From the existence of a $c=1$ plateau 
\begin{eqnarray}
(v/T)^{(\textrm{columnar})}_{\mathrm{c}} 
\leq
v/T 
\leq 
(v/T)^{(\textrm{staggered})}_{\mathrm{c}}
\label{eq: regime of criticality if v perturbation}
\end{eqnarray}
that extends to the left and right
of the Kasteleyn fixed point $v/T,T/U\to0$, 
we infer that criticality is preserved upon introducing a small but finite 
coupling $v/T$. 
Moreover, the behavior of the scaling dimensions presented in 
Fig.~\ref{fig: dimer Nc} (Bottom) tells us that the stiffness $K$ in the 
critical theory~(\ref{eq: c=1 action a}) varies continuously along the line 
of critical points.
Therefore we conclude that the introduction of the coupling $v$ realizes 
scenario III. 

The transition at $T=T^{(\textrm{columnar})}_{\mathrm{c}}$ has been
characterized by Alet et al. The quasi-long-range ordered phase%
~(\ref{eq: regime of criticality if v perturbation}) undergoes 
a Kosterliz-Thouless transition 
at $(v/T)^{(\textrm{columnar})}_{\mathrm{c}}$ 
to an ordered phase. This phase is characterized by an
alignment of parallel dimers along columns or rows and is therefore
called the columnar phase. 

On the other side, for $v>0$, our results show that the
quasi-long-range ordered phase~(\ref{eq: regime of criticality if v
perturbation}) terminates at $T=T^{(\textrm{staggered})}_{\mathrm{c}}$
where it undergoes a first-order phase transition to an ordered phase.
This phase is characterized by a staggering of parallel dimers along
two consecutive columns or rows and is therefore called the staggered
phase.

The phase diagram in the regime~(\ref{eq: regime 3})
has the same topology as the one for the 
constrained Ising model in the presence of a nearest-neighbor interaction 
(Fig.~\ref{fig: simple case}). 
%
%

\section{\label{sec: Quantum-discussion}
Quantum dominated regime, $|g|\ll |\Gamma^{\ }_{\mathrm{eff}}|$
        }
In Secs.%
~\ref{sec: Competing characteristic energy scales}-%
-\ref{sec: The square lattice dimer model}
we assumed that
$|\Gamma^{\ }_{\mathrm{eff}}|$ was the smallest energy scale in the problem. 
This is to be expected whenever 
$\Gamma^{\ }_{\mathrm{eff}} = \Gamma (\Gamma/U)^{n-1}$ 
is highly suppressed because of the order $n$ needed for virtual 
processes to return to an allowed configuration. 
Even if we initially had 
$\vert g \vert < \vert \Gamma \vert$, it is likely that 
$\vert\Gamma^{\ }_{\mathrm{eff}}\vert \ll \vert g \vert < \vert\Gamma\vert$, 
since $\vert \Gamma \vert \ll U$. 
However if the coupling constant $|g|$ 
for those terms that commute with the constraint is small compared to
$|\Gamma^{\ }_{\mathrm{eff}}|$, or in particular, if $g=0$, the discussion
above must be revisted. 

Let us consider here the case $g=0$ for simplicity. 
Independently of whether $\Gamma^{\ }_{\mathrm{eff}}$ is suppressed with 
respect to $\Gamma$ or not, there are now only three regimes of 
temperatures: 
\begin{subequations}
\label{eq: 3 regimes quantum}
\begin{eqnarray}
&&
T \ll 
|\Gamma^{\ }_{\mathrm{eff}}| \ll
U, 
\label{eq: regime AQ}
\\
&&
|\Gamma^{\ }_{\mathrm{eff}}|\ll T \ll U, 
\label{eq: regime BQ}
\\
&&
|\Gamma^{\ }_{\mathrm{eff}}| \ll
U \ll 
T.
\label{eq: regime CQ}
\end{eqnarray}
\end{subequations}

The constraint, as well as all other energy scales in the system, 
becomes negligible in regime ~(\ref{eq: regime CQ}) and the resulting 
physics is that of a featureless high-temperature phase, say
a paramagnetic phase if the degrees of freedom are exclusively magnetic.

As the temperature is lowered down to regime~(\ref{eq: regime BQ}), 
the system is still classical (at least from the point of view of its 
thermodynamic properties) but the constraint is now enforced. 
The physics is controlled by the proximity to the 
\textit{constrained entropic critical point} at 
$|\Gamma^{\ }_{\mathrm{eff}}|/T,T/U\to 0$.

Finally, in regime~(\ref{eq: regime AQ}) the system is fully quantum 
mechanical. 

\begin{figure}[!ht]
\center
\includegraphics[width=0.95\columnwidth]{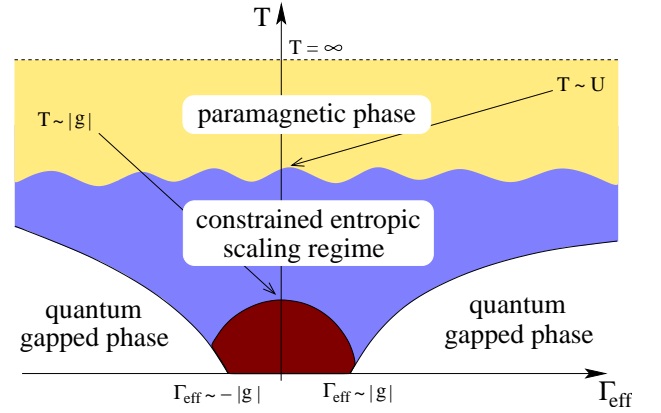}
\caption{
\label{fig: general ph diag b} 
(Color Online) -- 
Generic phase diagram for a strongly constrained quantum system that
satisfies conditions 1) to 6) from Sec.~\ref{sec: Introduction}. 
Parameter space encodes the competition between two energy scales,
the temperature $T$ and a characteristic energy scale
$\Gamma^{\ }_{\mathrm{eff}}$ given by the effective coupling of the 
off-diagonal term in the Hamiltonian, while $g$ and $U$ are held fixed. 
Here we chose to represent the case of a system exhibiting a zero 
temperature phase transition at a finite value of the ratio 
$\Gamma^{\ }_{\mathrm{eff}} / g$. 
A quantum critical scaling regime, if it exists, is restricted to
a region represented by the upper half of a disk of radius
$\Gamma^{\ }_{\mathrm{eff}}$ and centered at the origin
$(g,T)=(0,0)$ of parameter space. 
As in the case of Fig.~\ref{fig: general ph diag}, the system exhibits 
a \textit{constrained entropic scaling regime} 
at temperatures intermediate between the small characteristic quantum 
energy scale $\Gamma^{\ }_{\mathrm{eff}}$ and the large 
characteristic energy scale $U$ set by a strong constraint. 
At a fixed temperature, the constrained entropic scaling regime
terminates in a phase transition that needs not be continuous
upon increasing $|\Gamma^{\ }_{\textrm{eff}}|$. 
At fixed $\Gamma^{\ }_{\textrm{eff}}$, the constrained entropic scaling 
regime crosses over to the conventional high-temperature phase, 
say a paramagnetic one for spin degrees of freedom, 
when $T$ is of the order of $U$.
        }
\end{figure}

When $|g|$ is finite but much smaller than $|\Gamma^{\ }_{\mathrm{eff}}|$, 
these arguments should still be valid. 
Therefore, we can draw a qualitative phase diagram for a generic strongly 
constrained quantum system at fixed $g$ as a function of 
$\Gamma^{\ }_{\mathrm{eff}}$. 
In Fig.~\ref{fig: general ph diag b} we represent the phase diagram of a 
system exhibiting a zero temperature phase transition at a finite value of 
the ratio $\Gamma^{\ }_{\mathrm{eff}} / g$. 
Notice that the shaded dome around the origin or parameter space contains 
both the classical ordered phase that onsets when $|g|$ becomes larger than 
$|\Gamma^{\ }_{\textrm{eff}}|$, and the quantum critical scaling regime 
that is expected to appear in a cone-shaped region above the zero temperature 
quantum critical point. 
The actual details of this shaded region are highly system specific. 

It is worth mentioning that the existence of a zero temperature phase 
transition at $\Gamma^{\ }_{\textrm{eff}} \sim \pm |g|$ is by no means a 
necessity. 
It is also possible -- for example -- 
that the coupling $\Gamma^{\ }_{\mathrm{eff}}$ 
never destroys the ordered phase determined by the diagonal coupling 
$g$ at zero temperature, e.g., if it favors the same type of order. 
In this case there would be no quantum gapped phase in 
Fig.~\ref{fig: general ph diag b}, and the dome above the origin of 
parameter space would stretch out from $(-\infty,0)$ to $(\infty,0)$, 
without any phase transition at $(\pm \infty,0)$. 
The phase under the dome would be uniform and determined by the (fixed) 
value of $g$. 
This is so, for example, in the quantum 
Hamiltonian~(\ref{eq: example 4 bis eff}) for any fixed value of $g<0$, 
where the columnar ordered phase is stable at all values of 
$\Gamma^{\ }_{\mathrm{eff}}$ according to the numerical results in 
Ref.~\onlinecite{Syljuasen2005}. 
%
%

\section{\label{sec: Conclusions}
Conclusions
        }
In this paper we have presented a mechanism that leads to a large
temperature regime where critical scaling behavior appears as a
consequence of having a large energy scale $U$ in the problem. Hard
constraints are imposed on the system in the limit $U\to \infty$,
that project the original Hilbert space onto a space of allowed
configurations satisfying the constraints. Critical behavior in the
$U\to \infty$ limit occurs when correlations functions, calculated as
uniform averages over all of these projected states, are algebraically
decaying in space. We call the limit $U\to\infty$ 
followed by $T\to\infty$ at which criticality emerges
a \textit{constrained entropic critical point}. For large but finite
$U$, and in the presence of other couplings $|g|,|\Gamma| \ll U$ in
the problem, the physics at temperatures large compared to all other
couplings but small compared to $U$ 
is still controlled by the proximity to the constrained
entropic critical point. Indeed, this hierarchy among the
couplings opens a rather 
large window of temperatures for which a 
constrained entropic critical regime
exists as a consequence of the proximity to the 
constrained entropic critical point.
We expect the constrained entropic critical regime 
to be qualitatively different in general 
from the quantum critical regime associated 
to a putative quantum critical point at zero temperature. 
The picture is summarized in Fig.~\ref{fig: general ph diag}.

The mechanism for criticality at high temperature
discussed here should be applicable to some, but not all,
problems with a large dominant energy scale. Examples can
be found in frustrated magnets, say pyrochlore antiferromagnets
where a magnetic field that induces a magnetization
plateau at half the value of full magnetization
supplies a strong constraint.%
\cite{Penc04,Bergman05a,Bergman05b}
Furthermore, since the order $n$ of a virtual
process leading to the effective kinetic quantum term 
$
\Gamma^{\ }_{\mathrm{eff}}
= 
\Gamma (\Gamma/U)^{n-1} 
$ 
is usually large (it is of order 9 in Ref.~\onlinecite{Bergman05a}), 
the window of temperatures for which the system is in the
quantum regime can be extremely small 
($T < |\Gamma^{\ }_{\mathrm{eff}}| \ll |\Gamma|$). 
Hence, in practice, even for
the smallest temperatures accessible experimentally, 
these constrained systems should either order or be
within the constrained entropic critical regime. 
What determines whether the system orders or not is the presence 
of another energy scale $g$, set by the coupling strengths 
of additional terms in the Hamiltonian that commute 
with the $U$-term imposing the constraint.

Another example for which the idea of high-temperature criticality may
apply is that of the fluctuations about a so-called $d$-density wave
(DDW).\cite{Chakravarty01} In this instance, the order may form
locally at some large energy $\Delta^{\ }_d$, but the current loop
directions may fluctuate.~\cite{Chakravarty02,Syljuasen05} At low
temperatures, the system should order in a given current pattern.
Above the global ordering temperature, this system should display
high-energy constrained entropic criticality, because it resembles an
ice model as long as $T$ is below $\Delta^{\ }_d$.

In the single-band Hubbard model, 
with a strong local on site repulsion only, there is \textit{no} 
constrained entropic critical point in the $U\to\infty$ limit. 
It has been recently proposed by Philip Phillips and co-workers that
the order in which the limit $U\to\infty$ and the 
thermodynamic limit $L\to\infty$ ($L$ being the linear size of the system) 
are taken has implications for hole transport in the single-band Hubbard 
model.\cite{Phillips04}
The issue is that, for finite $U$, there is a characteristic
length scale associated with the distance between doubly occupied
sites, and hence for system sizes greater than this distance one has a
finite density of such ``defects''. The appearance of this extra
length scale was suggested as a way to resolve the issue of the
breakdown of the one-parameter scaling picture for quantum criticality in
the cuprates.\cite{Phillips05} 
While we do not know how to connect their results to ours, 
it seems that there is one common theme, that high-energy terms
find their way to affect the physics at intermediate temperatures.

The chances to find a constrained entropic critical point improve if
one adds nearest-neighbor, next-nearest neighbor, etc..., couplings,
i.e., in extended versions of the Hubbard model, at some commensurate
fillings. For example, the presence of very strong nearest-neighbor
interactions at 1/4 filling leads to a classical checkerboard
configuration, where one of the two sublattices of the bipartite
square lattice is filled and the other empty. 
If one now starts changing the doping away from 1/4
filling, extra holes or particles will tend to cluster in
stripes.%
\cite{Low1994,Mila1994,Henley2001,McKenzie2001,Arrigoni2002,Zhang2003, 
Arrigoni2004} 
Other longer range couplings that
commute with the constraint could lead to charge ordering, perhaps in
the form of static stripes or other patterns at low temperatures. Now,
at high temperatures, when charge order is destroyed, these stripes
meander and fluctuate, and the system could be in an entropic critical
regime. Certainly, these are speculative thoughts, in contrast to the
concrete cases of the constrained quantum Heisenberg and transverse
field Ising models that we presented as examples of constrained
entropic criticality. But the discussion suggests how it is not 
implausible that a constrained entropic critical point may play a role
in the physics of the extended Hubbard model near commensurate
fillings.
%
%

\section*{Acknowledgments}

We would like to thank S. Chakravarty, V. Dotsenko, and M. Picco for
their insightful comments. We are particularly indebted to E. Fradkin
for several useful discussions, and for his careful reading of the
manuscript. This work is supported in part by the NSF Grants
DMR-0305482 and DMR-0403997 (CC and CC).
%
%

\end{document}